\documentclass[aps,prd,reprint,twocolumn,preprintnumbers,floatfix,nofootinbib]{revtex4-1}
\usepackage[utf8]{inputenc}
\usepackage[dvips]{graphicx}
\usepackage[dvipsnames]{xcolor}
\usepackage{color}
\usepackage{relsize}
\usepackage{float}
\usepackage{graphics}
\usepackage{epstopdf}
\usepackage{hyperref}
\usepackage{mathrsfs}
\usepackage{amssymb}
\usepackage{physics}
\usepackage{booktabs}

\usepackage[normalem]{ ulem }
\usepackage{amsthm}
\usepackage{amsmath}
\usepackage{cancel}
\usepackage{fontawesome} % for code link icons  
\usepackage[caption=false]{subfig}
\usepackage{tabularx,ragged2e} % new
\usepackage{lipsum}
\newcolumntype{L}{>{\RaggedRight\arraybackslash}X}
\newcommand{\equaref}[1]{Eq.~(\ref{#1})}
\usepackage{hyperref}

\newcommand{\be}{\begin{equation}}
	\newcommand{\ee}{\end{equation}}
\newcommand{\bea}{\begin{eqnarray}}
	\newcommand{\eea}{\end{eqnarray}}
\newcommand{\ba}{\begin{aligned}}
	\newcommand{\ea}{\end{aligned}}

\newcommand{\figref}[1]{Fig.~\ref{#1}}

\newcommand{\fd}[2]{\frac{d #1}{d #2}}
\newcommand{\secref}[1]{Sec.~\ref{#1}}

\newcommand{\tabref}[1]{Table~\ref{#1}}

\newcommand\myshade{80}
\colorlet{mylinkcolor}{ForestGreen}
\colorlet{mycitecolor}{Red}
\colorlet{myurlcolor}{violet}

\hypersetup{
	linkcolor  = mylinkcolor!\myshade!black,
	citecolor  = mycitecolor!\myshade!black,
	urlcolor   = myurlcolor!\myshade!black,
	colorlinks = true
}

\usepackage{pgfplots}
\renewcommand{\equaref}[1]{Eq.~\eqref{#1}}

\pgfplotsset{compat=1.17}

\begin{document}
	\sloppy  
	
	\preprint{IPPP/23/20}
	
	\vspace*{1mm}
	
	\title{
		JUNO as a Probe of the Pseudo-Dirac Nature using Solar Neutrinos
	}
	
	\author{Jack Franklin}
	\email{jack.d.franklin@durham.ac.uk}
	\author{Yuber F. Perez-Gonzalez}
	\email{yuber.f.perez-gonzalez@durham.ac.uk}
	\author{Jessica Turner}
	\email{jessica.turner@durham.ac.uk}
	
	\affiliation{Institute for Particle Physics Phenomenology, Durham University, South Road DH1 3LE, Durham, U.K.}
	
	\begin{abstract}
		{
			It remains a possibility that neutrinos are pseudo-Dirac states, such that a generation is composed of two maximally mixed Majorana neutrinos separated by a very small mass difference. We explore the physics potential of the JUNO experiment in constraining this possibility using the measurement of solar neutrinos. In particular, we investigate cases where one or three sterile states are present in addition to the active states.
			We consider two scenarios: one where JUNO's energy threshold allows for the measurement of $pp$ solar neutrinos, and the case where JUNO can only measure $^7$Be neutrinos and above.
			We find that JUNO will be able to constrain pseudo-Dirac mass splittings of $\delta m^2 \gtrsim 2.9\times 10^{-13}~{\rm eV^2}$ for the scenario including $pp$ solar neutrinos, and $\delta m^2 \gtrsim 1.9\times 10^{-12}~{\rm eV^2}$ when the measurement only considers $^7$Be monochromatic neutrinos, at the $3\sigma$ C.L.
			Thus, including $pp$ neutrinos will be crucial for JUNO to improve current constraints on the pseudo-Dirac scenario from solar neutrinos.
		}
	\end{abstract}

	\maketitle
	
	\section{Introduction}
	It has been established for more than two decades that neutrinos have small but non-zero masses, and 
	new degrees of freedom beyond the Standard Model (SM) must exist to accommodate this observation \cite{Super-Kamiokande:1998kpq,SNO:2002tuh}. The seesaw mechanisms, \cite{Mohapatra:1979ia,Gell-Mann:1979vob,Yanagida:1979as,Minkowski:1977sc,Mohapatra:1980yp,Magg:1980ut,Lazarides:1980nt,Wetterich:1981bx,Foot:1988aq,Ma:1998dn} introduce heavy states that mediate the Weinberg operator and generate light neutrino masses after electroweak symmetry breaking. Purely Dirac neutrino masses also require introducing new SM gauge-singlet fermionic degrees of
	freedom. While technically natural, this hypothesis is often considered contrived as it requires extremely small Yukawa couplings. 
	The key distinction between these two possibilities is that the former mechanism requires that lepton number is violated while the latter does not. However, lepton number may be violated such that neutrinos are still almost Dirac particles, i.e. ``pseudo-Dirac'' neutrinos
	\cite{Wolfenstein:1981kw,Petcov:1982ya,Bilenky:1983wt,Foot:1995pa,Chang:1999pb}. 
	The small Majorana masses lift the degeneracy of mass-eigenvalues, resulting in almost degenerate pairs of eigenstates with tiny mass splittings. 
	Such small breaking of the lepton number could be of gravitational origin since quantum gravity effects are expected to break global symmetries. 
	Thus, the dimension-5 Weinberg operator could be Planck-suppressed, generating tiny Majorana masses~\cite{Carloni:2022cqz}.
	Moreover, models that predict light Dirac neutrinos ---see e.g.~Refs~\cite{Mohapatra:1987hh,Babu:1988yq,Farzan:2012sa,Ma:2016mwh,Saad:2019bqf,Jana:2019mez,Babu:2022ikf}--- would typically predict pseudo-Dirac neutrinos after the inclusion of higher-dimensional operators which are suppressed by the Planck scale.
	Likewise, if Dirac neutrino masses come from the spontaneous breaking of a gauge symmetry, such as left-right symmetric theories~\cite{Davidson:1987mh, Balaji:2001fi, Borah:2017leo}, additional effects could generate a small Majorana mass terms, leading to pseudo-Dirac neutrinos~\cite{Carloni:2022cqz}. 
	
	From the experimental perspective, the pseudo-Dirac nature's determination mainly consists of searching for active-sterile oscillations driven by an additional mass splitting, $\delta m_k^2$.
	Other signatures of lepton number breaking, especially the measurement for neutrinoless double beta decay, would be highly suppressed due to the smallness of the Majorana mass term.
	Current solar experiments constrain the mass splitting $\delta m_k^2 \lesssim 10^{-12}$~eV$^2$~\cite{Chen:2022bxn, Ansarifard:2023} finding a slight preference for a non-zero mass splitting of $1.5\times 10^{-11}~{\rm eV^2}$~\cite{Ansarifard:2023}, while, using $pp$ neutrinos, the future DARWIN dark matter detector could be sensitive to values of $\delta m_k^2 \sim 10^{-13}~{\rm eV^2}$~\cite{deGouvea:2021ymm}.
	The bound is much weaker for atmospheric neutrinos,  $\delta m_k^2 \lesssim 10^{-4}$~eV$^2$~\cite{Beacom:2003eu}.
	Since the mass splitting can be arbitrarily small, neutrinos travelling astrophysical distances can place the most stringent limits on this scenario.
	The analysis of the SN1987A neutrinos can exclude values between $[2.55, 3.01]\times 10^{-20}~{\rm eV^2}$~\cite{Martinez-Soler:2021unz}; meanwhile, tiny values of $\delta m_{k}^2 \sim 10^{-24}~{\rm eV^2}$ could be tested by measuring the diffuse supernova neutrino background (DSNB)~\cite{DeGouvea:2020ang}.
	High energy neutrinos only can explore larger values of the mass splittings, $10^{-18}~{\rm eV^2}\lesssim \delta m^2 \lesssim 10^{-12}~{\rm eV^2}$, due to their high boost~\cite{Beacom:2003eu, Rink:2022nvw,Carloni:2022cqz}.
	
	In general, we would require low-energy neutrinos travelling large distances to observe active-sterile oscillations for tiny mass splittings. 
	As mentioned, supernova neutrinos could help constrain values as small as $10^{-20}~{\rm eV^2}$.
	However, the occurrence of a supernova explosion is somewhat uncertain. 
	At the same time, although the DSNB is a guaranteed flux, its smallness combined with the large backgrounds plague its search, which means that competitive constraint will only be obtained a decade after its discovery.
	Until then, solar neutrinos offer an alternative to measure pseudo-Dirac oscillations.
	Low energy fluxes, such as $pp$ or $^7$Be neutrinos from the p-p reaction chain, will be measured with high precision in the next generation of experiments, thus leading to possible improvements on the limits of the pseudo-Dirac scenario.
	In this context, the Jiangmen Underground Neutrino Observatory (JUNO)~\cite{JUNO:2015zny} offers an additional facility to constrain the presence of additional active-sterile oscillations on top of the standard ones.
	JUNO is expected to have an energy resolution of 3$\%\sqrt{E_r/\mathrm{MeV}}$, and an energy threshold that could be of order $E_{r}\sim {\cal O}(200)~{\rm keV}$~\cite{JUNO:2015zny}, such that it could measure the intermediate-energy --$^7$Be, CNO-- solar neutrino fluxes at the $\sim 10\,\%$ level after six years of data taking, depending on the extent to which the backgrounds can be placed under control~\cite{JUNO:2023zty}.
	However, the situation is still unclear for the lower energetic $pp$ neutrinos since the $^{14}$C and $^{14}$C-pile up background should be dominant for energies $E_r\lesssim 160$ keV.
	With this in mind, in this paper, we determine the sensitivity of JUNO in constraining active-sterile oscillations in the pseudo-Dirac framework.
	We consider three possibilities: the first one where only intermediate-energy neutrinos, specifically $^7$Be neutrinos, are measured, a second more optimistic scenario where the higher energy tail of $pp$ neutrinos, above the threshold mentioned above of $E_r\lesssim 160$ keV, are included in the measurement and finally a third intermediate scenario where it is still possible to detect the tail of the $pp$ neutrino spectrum, but $^7$Be neutrinos overwhelmingly dominate the signal.
	We find that measurement of the $pp$ neutrinos is crucial for JUNO to improve current constraints on the mass splitting.
	
	The paper is organised as follows. First, in Sec.~\ref{sec:PDosc}, we consider the generalities of pseudo-Dirac neutrino oscillations in vacuum and in matter since these will be relevant for discussing the modifications that can appear in solar neutrinos. 
	We also review approximated formul\ae~for the specific case of solar $pp$ neutrinos and describe the numerical approach we have implemented for $^7$Be neutrinos.
	We consider solar neutrinos in Sec.~\ref{sec:sol} and review their measurement in the JUNO experiment using neutrino-electron scattering.
	Sec.~\ref{sec:osc} describes the statistical procedure that we have implemented for the analysis of active-sterile oscillations, and we present our results in Sec.~\ref{sec:res}.
	Finally, we draw our conclusions in Sec.~\ref{sec:conc}.
	Throughout this manuscript, we use natural units where $\hbar = c = k_{\rm B} = 1$.

	%%%%%%%%%%%%%%%%%%%%%%%%%%%%%%%%%%%%%%%%%%%%%%%%%%%%%%%%%%%%%%%%%%%%%%%%%%%%%%%%
	\section{Pseudo-Dirac Neutrino Oscillations}\label{sec:PDosc}
	
	One of the open problems in particle physics is understanding the origin of neutrino masses. The most straightforward extension of the SM to address this would be to include right-handed neutrino fields $N_{R}^i$ and implement the Higgs mechanism to generate neutrino masses as done for the other charged fermion.
	This simple approach implies a dilemma:  a na\"ive estimate would indicate that Yukawa couplings for neutrinos need to be extremely small, ${\cal O} (10^{-12})$, to produce masses of $m_\nu\sim {\cal O}({\rm eV})$.
	Thus, one may wonder if there is a way to understand the smallness of such couplings.
	Let us note, nevertheless, that the additional right-handed neutrinos would be singlets of the SM.
	Therefore, gauge invariance does not forbid the presence of Majorana mass terms for such degrees of freedom.
	The most general mass Lagrangian for neutrinos would then be
	\begin{align}
		\mathscr{L}_{\nu}  = -Y_{\alpha i} \overline{L_\alpha} \widetilde{H} N_{R}^i + \frac{1}{2}\overline{(N_{R}^i)^c} M_R^{ij} N_{R}^j\,,
	\end{align}
	where $L_\alpha$, $\widetilde{H}$, are the SM left-handed lepton and conjugate Higgs doublets, $Y_{\alpha i}$ the Yukawa matrix, $M_R^{ij}$ the Majorana mass matrices, respectively, and the $c$ superscript on the Majorana mass term indicates charge conjugation. After the electroweak symmetry breaking, the neutrino mass Lagrangian can be rewritten in a simpler form,
	\begin{align}
		\mathscr{L}_{\nu}  = - \frac{1}{2} \overline{\psi^c} M \psi\,,
	\end{align}
	where
	\begin{align}
		\psi = \begin{pmatrix}
			\nu_L\\
			(N_R)^c
		\end{pmatrix},\quad
		M    = \begin{pmatrix}
			0_3 & Y v/\sqrt{2}\\            
			Y v/\sqrt{2} & M_R
		\end{pmatrix}\,,
	\end{align}
	with  $v$ the Higgs vacuum expectation value (vev), $\nu_L = (\nu_e, \nu_\mu, \nu_\tau)^T$ and $N_R=(N_R^1, N_R^2, \ldots)^T$ vectors for the left- and right-handed neutrino fields. At this point, we have not specified any hierarchy between the Higgs vev and the scale of the Majorana mass matrix $M_R$.
	The renowned seesaw mechanism~\cite{Mohapatra:1979ia,Gell-Mann:1979vob,Yanagida:1979as,Minkowski:1977sc,Mohapatra:1980yp,Magg:1980ut,Lazarides:1980nt,Wetterich:1981bx,Foot:1988aq,Ma:1998dn} establishes that if a large hierarchy between the scales $M_R\gg Y v$ exists, the neutrino masses will be suppressed by a factor of $m_\nu \propto Y^T (M_R)^{-1} Y v^2$ with respect to the electroweak scale.
	Such a scenario has attracted much attention since it would also explain the observed matter-antimatter asymmetry in the Universe~\cite{Yanagida:1979as}.
	There is, however, the possibility that the Majorana mass scale is suppressed with respect to the electroweak scale, $M_R\ll Y v$, if, for instance, such Majorana mass terms are Planck-suppressed.
	In this scenario, which we will denote as \emph{pseudo-Dirac}\footnote{In the literature, this scenario is also denoted as \emph{quasi-Dirac}.}, the accidental lepton number conservation is softly broken by the Majorana mass term $M_R$. Hence, it lifts the mass degeneracy in a Dirac neutrino between its left- and right-handed components.
	Crucially, in this scenario, neutrinos behave mostly as Dirac particles to such a degree that lepton-number violation processes will be highly suppressed, making an experimental discovery via lepton-violating processes difficult.
	
\begin{figure}[t]
		\centering
		\includegraphics[height=0.3\textwidth]{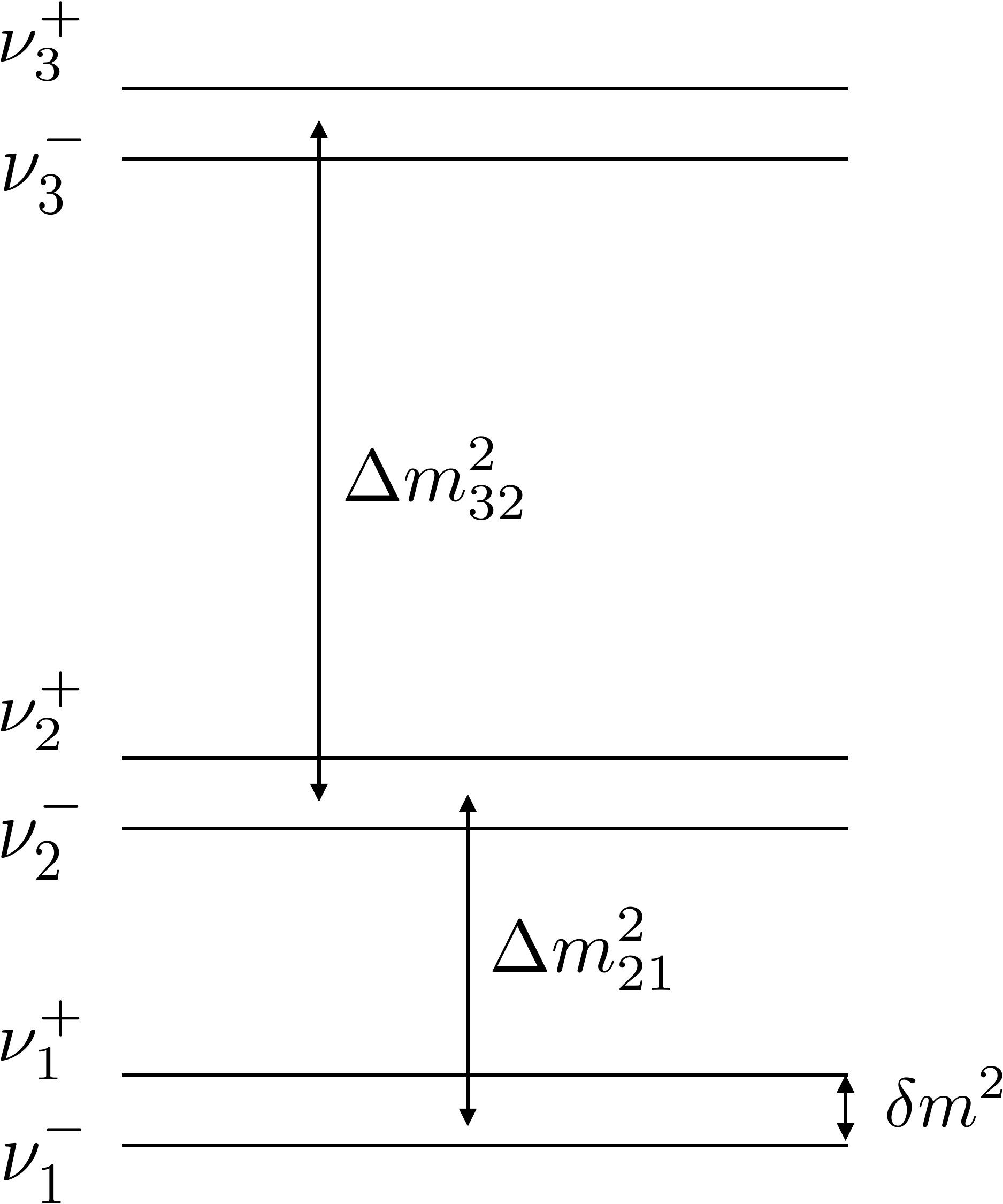}
		\caption{ The neutrino mass spectrum, showing the usual solar
			and atmospheric mass differences, as well as the pseudo-Dirac
			splittings in each generation, $\delta m^2$. In principle, the PD mass squared splitting can differ between generations but for simplicity, we assume it is the same between each generation. The mass eigenstates are denoted as $\nu^+_i$ and $\nu^-_i$ for PD pair $i$. The active ($\nu_a$) and sterile ($\nu_s$) components of each pseudo-Dirac pair are a maximal mixtures of the mass eigenstates}
		\label{fig:PDn}
	\end{figure}

	Nevertheless, the pseudo-Dirac scenario predicts oscillations between the active and sterile components, which leads to modifications of the standard oscillations, especially those for neutrinos travelling long distances.
	Thus, these active-sterile oscillations can lead to observable effects in different facilities. 
	Since our work will focus on the constraints that JUNO could place on pseudo-Dirac neutrinos from solar neutrinos, let us describe the mass spectrum and its mixing in the scenario where we add three right-handed neutrinos.
	Let us first consider the general case in which we do not consider a specific hierarchy between the Majorana mass matrix and the electroweak scale.
	The mass matrix, $M$, will be diagonalised by a $6\times 6$ unitary matrix, $\mathscr{V}$, obtained by multiplying 15 complex rotation matrices~\cite{Anamiati:2019maf}.
	For simplicity, we will assume that mixing exists only between the pseudo-Dirac pairs, which we will label $1-4$, $2-5$ and $3-6$, see the diagram in Fig.~\ref{fig:PDn}. 
	Thus, we consider the non-zero mixing angles $\theta_{14}, \theta_{25}, \theta_{36}$, so that the mixing matrix will be
	\begin{align}
		\mathscr{V} = U_{23} U_{13} U_{12} U_{14} U_{25} U_{36}\,.
	\end{align}
	From the definition of the mixing matrix $\mathscr{V}$, we can define mass eigenstates $\nu_{i}^\pm$, with definite masses $m_{i}^\pm$, as usual~\cite{Kobayashi:2000md}
	\begin{align*}
		\psi = \mathscr{V}\cdot
		\begin{pmatrix}
			\nu_{i}^+\\
			\nu_{i}^-
		\end{pmatrix}.
	\end{align*}
	In the limit in which the lepton number breaking scale is $M_R\ll Y v$, the mixing between the mass eigenstates becomes maximal in such a way that $\theta_{14}=\theta_{25}=\theta_{36}=\pi/4$, and the mixing matrix $\mathscr{V}$ can be parametrised as~\cite{Kobayashi:2000md}
	\begin{align*}
		\mathscr{V} = \begin{pmatrix}
			U^{3f} & 0 \\
			0      & U_R
		\end{pmatrix} \cdot 
		\frac{1}{\sqrt{2}}
		\begin{pmatrix}
			1_3 & i 1_3\\
			\varphi & -i \varphi
		\end{pmatrix}\,,
	\end{align*}
	where $U^{3f}$ is the standard Pontecorvo-Maki-Nakagawa-Sakata (PMNS) matrix for three flavour oscillation, and $\varphi = {\rm diag}(e^{-i\phi_1},e^{-i\phi_2},e^{-i\phi_3})$ is a matrix containing arbitrary phases, and $1_3$ denotes the $3 \times 3$ identity matrix. The neutrino fields, in the flavour basis, take a simpler form in the pseudo-Dirac limit,
	\begin{align}
		\nu_{\alpha} = \frac{U_{\alpha k}^{3f}}{\sqrt{2}} (\nu_k^+ + i\nu_k^-)\,.
	\end{align}
	From this, we observe that a flavour eigenstate is a maximally-mixed superposition of two mass eigenstates with almost degenerate masses, $m_{k,\pm}^2=m_k^2\pm \delta m_k^2/2$.
	Current limits indicate that the mass splittings $\delta m _k^2$ coming from soft lepton number breaking must be much smaller than the solar and atmospheric mass differences.
	%%%%%%%%%%%%%%%%%%%%%%%%%%%%%%%%%%%%%%%%%%%%%%%%%%%%%%%%%
	%%%%%%%%%%%%%%%%%%%%%%%%%%%%%%%%%%%%%%%%%%%%%%%%%%%%%%%%%
	
	In general, neutrino evolution in the flavour basis will be dictated by the usual Schr\"odinger-like equation
	\begin{align}
		i\frac{d}{dt} \psi = H \psi,
	\end{align}
	where the Hamiltonian, including matter effects, is~\cite{Kobayashi:2000md}
	\begin{align}
		H = \frac{1}{2E_\nu}\left[\mathscr{V}M_{\rm diag}\mathscr{V}^\dagger + \mathscr{A}\right],
	\end{align}
	with $E_\nu$ the neutrino energy, $M_{\rm diag}$ the $6 \times 6$ diagonal mass matrix, and $\mathscr{A}$ the matter potential 
	\begin{align}
		\mathscr{A} = \sqrt{2}G_F E\, {\rm diag}\left(2N_e-N_n,-N_n,-N_n, 0, 0, 0 \right),
	\end{align}
	where $N_e$ and $N_n$ are the electron and neutron number density, respectively.
	In the scenario where the pseudo-Dirac mass splittings are much smaller than the solar and atmospheric ones, $\delta m^2 \ll \Delta m_{21, 31}^2$, matter effects will only affect the propagation of the pseudo-Dirac pairs, in a similar fashion to the standard MSW.
	Once the evolution equation is solved, we can obtain neutrino oscillation probabilities in a standard manner. 
	The numerical approach will be specified later on.
	
	As a succinct reminder for the reader, let us briefly recall the properties of Solar neutrinos.
	Electron neutrinos are produced in the Sun's core via two distinct processes, the p-p chain and the CNO cycle, where the latter is subdominant.
	The largest number of neutrinos, $\sim 91\%$ of the total, is produced from p-p reactions producing deuterium, $p^++p^+\longrightarrow d + e^+ + \nu_e$. 
	These $pp$ neutrinos have a broad spectrum with a maximum energy of $\sim 420$~keV.
	Other nuclear interactions belonging to the p-p chain produce neutrinos in smaller amounts, such as $^7$Be ($\sim 7.3\%$), $pep$ ($\sim 0.2\%$), $^8$B ($0.01\%$) and $hep$ ($\sim 1.4\times 10^{-5}\%$); the CNO chain produces the remainder of neutrinos $\sim 1.5\%$.
	For our purposes, we will focus in the two most abundant types of neutrinos, $pp$ and $^7$Be. 
	Their specific spectrum will be presented in the next section.
	
	We are interested in low and intermediate-energy solar neutrinos, so we can consider analytical approximations to the oscillation probabilities~\cite{deGouvea:2021ymm}.
	For $pp$ neutrinos, which have energy $E_\nu \lesssim 420$ keV, the standard matter effects are negligible, and thus we can approximate the mixing as modifying the standard solar oscillation probabilities by including active-sterile oscillation of each pair. At these energies, the standard vacuum oscillations average out the distance dependence factors due to the large production region leaving only powers of the PMNS matrix elements. This simplifies the form of the oscillation probabilities:
	\begin{subequations}\label{eq:solar_probs}
		\begin{align}
			P_{ee} &= 
			\abs{U^{3f}_{e1}}^4 P^{2f}_{ee}(\theta_{14}, \delta m^2_1) + \abs{U^{3f}_{e2}}^4 P^{2f}_{ee}(\theta_{25}, \delta m^2_2)\notag\\ 
			&+ \abs{U^{3f}_{e3}}^4 P^{2f}_{ee}(\theta_{36}, \delta m^2_3) \,,\\
			P_{es} &= \abs{U^{3f}_{e1}}^2 \left(1 - P^{2f}_{ee}(\theta_{14}, \delta m^2_1)\right)\notag\ \\ 
			&+ \abs{U^{3f}_{e2}}^2 \left(1-P^{2f}_{ee}(\theta_{25}, \delta m^2_2)\right)\notag\\\ 
			&+ \abs{U^{3f}_{e3}}^2 \left(1-P^{2f}_{ee}(\theta_{36}, \delta m^2_3)\right)\,,\\
			P_{ea} &= 1 - P_{ee} - P_{es}\,.
		\end{align}
	\end{subequations}
	where $P_{ee}$ is the electron neutrino survival probability and $P_{es}$ ($P_{ea}$) the electron neutrino to sterile (other active flavour) oscillation probability.
	Separating the muon and tau flavour probabilities is unnecessary as they have identical contributions to the scattering cross section, as will be reviewed in \secref{sec:sol}. 
	We can further approximate the two-neutrino oscillation probabilities by analysing the matter effects on the evolution of each pseudo-Dirac pair, as follows.
	\\
	
	%%%%%%%%%%%%%%%%%%%%%%%%%%%%%%%
	\textbf{A.} $10^{-10} \mathrm{eV}^2 \lesssim \delta m^2 \lesssim 10^{-6} \mathrm{eV}^2$ ---
	Vacuum oscillations are averaged for mass splittings in this range, so we can remove any dependence on the Earth-Sun distance. However, in this regime, matter effects are important for the active-sterile oscillations, so the MSW effect must be considered. To do so, we must take into consideration the non-adiabaticity of the Solar density profile in this regime, which is done via the crossing probability $P_c$, in general, given by~\cite{Parke:1986jy,deGouvea:1999wg,Friedland:2000cp}
	\begin{equation}
		P_c = \frac{e^{-\gamma \sin^2\theta}-e^{-\gamma}}{1 - e^{-\gamma}}\,,
	\end{equation}
	where $\theta$ is the mixing angle between the two neutrino states, and the non-adiabaticity parameter is given by~\cite{deGouvea:1999wg,Friedland:2000cp}
	\begin{align}
		\gamma=2\pi r_0^k \frac{\delta m_k^2}{2 E_\nu}\,,
	\end{align}
	with $r_0^k$ a distance obtained by performing an exponential fit of the matter potential inside the Sun, $N_{ij}(r) = N_0^{ij} \exp(-r/r_0^{ij})$.
	Such a matter potential will depend on the specific pseudo-Dirac scenario to be tested~\cite{deGouvea:2021ymm},
	\begin{align}
		N_{ij}(r) = 
		\begin{cases}
			N_e(r) \cos^2\theta_{13} \cos^2\theta_{12}-\frac{1}{2}N_n(r) & ij = 14\\
			N_e(r) \cos^2\theta_{13} \sin^2\theta_{12}-\frac{1}{2}N_n(r) & ij = 25\\
			N_e(r) \sin^2\theta_{13} -\frac{1}{2}N_n(r) & ij = 36\\
		\end{cases}.
	\end{align}
	For our purposes, we consider the electron and neutron number densities predicted by the Solar Model AGSS09 from Ref.~\cite{Vinyoles:2017}.
	On the other hand, vacuum oscillations between the Sun and the Earth average out for these parameters.
	The active-sterile two-neutrino probability will then follow the Parke formula~\cite{Parke:1986jy}
	\begin{align}
		P^{2f}_{ee}(\theta_{ij}, \delta m^2_k) = \frac{1}{2}+\left(\frac{1}{2}- P_c\right)\cos^2 2\theta^m_{ij} \cos^2 2\theta_{ij}\,,
	\end{align}
	where $ij=\{14,25,36\}$, and the usual expression gives the effective mixing angle in the Sun
	\begin{equation}
		\cos2\theta^m_{ij} = \frac{\delta m_k^2 - 2 E_\nu N_0^k}{\sqrt{(\delta m_k^2 \cos2\theta_{ij} -2E_\nu N_0^k)^2 - (\delta m_k^2 \sin2\theta_{ij})^2}}.
	\end{equation}
	\\
	
	%%%%%%%%%%%%%%%%%%%%%%%%%%5
	\textbf{B.} $10^{-11} \mathrm{eV}^2\lesssim \delta m^2 \lesssim 10^{-10}\mathrm{eV}^2 $ ---
	The vacuum oscillations also play an important role in these ranges of pseudo-Dirac mass-splitting values.
	In this intermediate situation, we can recast the analytical approximations obtained for a two-flavour oscillation in Refs~\cite{deGouvea:1999wg,Friedland:2000cp} to 
	\begin{multline}
		P^{2f}_{ee}(\theta_{ij}, \delta m^2_k) = P^\prime_c \cos^2\theta_{ij} + (1-P^\prime_c)\sin^2\theta_{ij} \\ 
		- \sqrt{P_c(1-P_c)} \cos^2\theta_{ij}^m \sin^2\theta_{ij} \cos \left( \frac{\delta m_k^2 L_{\odot}}{2 E_\nu}\right)\,,
	\end{multline}
	where
	\begin{equation}
		P^\prime_c = P_c \sin^2\theta_{ij}^m + (1 - P_c) \cos^2\theta_{ij}^m\,,
	\end{equation}
	and $L_{\odot}$ is the Sun-Earth distance.
	\\
	
	%%%%%%%%%%%%%%%%%%%%%%%%%%%%
	\textbf{C.} $\delta m^2 \lesssim 10^{-11}\mathrm{eV}^2$ ---
	Finally, for mass splittings below $10^{-11}~{\rm eV^2}$, matter effects are not important for active-sterile oscillations. Thus, the two-neutrino probability will have the standard form in vacuum,
	\begin{equation}
		P^{2f}_{ee}(\theta_{ij}, \delta m^2_k) = 1 - \sin^2(2\theta_{ij}) \sin^2\left( \frac{\delta m_k^2 L_{\odot}}{4 E_\nu}\right).
	\end{equation}
	
	For the higher energy $^7$Be neutrino line, at 862 keV, the previous analytic approximations would lead to probabilities that do not reproduce the correct values due to matter effects affecting the active neutrinos. Thus, numerical calculations were performed using the \textit{slab approximation} method \cite{GiuntiKim}. This involves the discretisation of the matter density profile of the Sun into slabs of constant density with some length $\Delta x$, through which the propagation of the neutrino amplitude can be calculated. Thus, the amplitude of a neutrino after passing through a varying density profile can be approximated as
	\begin{equation}\label{eq:nu_amp}
		\mathcal{A} = \prod_{s=0}^{N} \mathscr{V}_s \exp\left(-i \frac{m^2_s \Delta x_s}{2 E_\nu}\right) \mathscr{V}^\dagger_s \mathcal{A}_0\,,
	\end{equation}
	where $\mathscr{V}_s$ and $m^2_s$ are the effective mixing matrix and effective mass squared difference matrix, respectively, in the slab $s$. These are obtained by diagonalising the Hamiltonian in the medium. The initial amplitude, $\mathcal{A}_0$, for solar neutrinos, which are produced as pure $\nu_e$ states, is $(1,0,0,0,0,0)^T$ in the flavour basis. We can obtain the probability from the amplitude:
	\begin{equation}
		P_{e\alpha}^{1\nu} = \left| \mathcal{A}_\alpha \right|^2\,.
	\end{equation}
	However, this is only for a single neutrino originating from one position. In reality, the Sun produces many neutrinos over a large region. Since we do not know where a detected neutrino was produced, we must average the probability over the entire production region yielding the Solar probability:
	\begin{equation}\label{eq:avg_prob}
		P_{e\alpha} = \int^{r_b}_{r_a} dr \rho(r) P^{1\nu}_{e\alpha}(r) \approx \sum_{i=0}^N \rho(r_i) P^{1\nu}_{e\alpha}(r_i)\,,
	\end{equation}
	where $\rho$ is the production probability as a function of the radial position in the Sun, and $P^{1\nu}_{e\alpha}$ now depends on where the neutrino was produced. For this work, we approximate $\rho$ as a window function between $0.02\,R_\odot$ and $0.125\,R_\odot$ for $^7$Be, which we found to be within a few percent of the probability calculated using theoretical predictions of $\rho$. This is a reasonable approximation as, qualitatively, the averaged probability depends primarily on the length over which it is averaged rather than the exact distribution of the production. Further, the production region for $^7$Be is highly concentrated over this region \cite{Lopes:2013} and so should be approximated well by a uniform distribution over this length.
	%\jt{Can you state why this is a reasonable approximation}
	
	%%%%%%%%%%%%%%%%%%%%%%%%%%%%%%%%%%%%%%%%%%%%%%%%%%%%%%%%%%%%5
	\begin{figure}[t]
		\centering
		\includegraphics[width=0.48\textwidth]{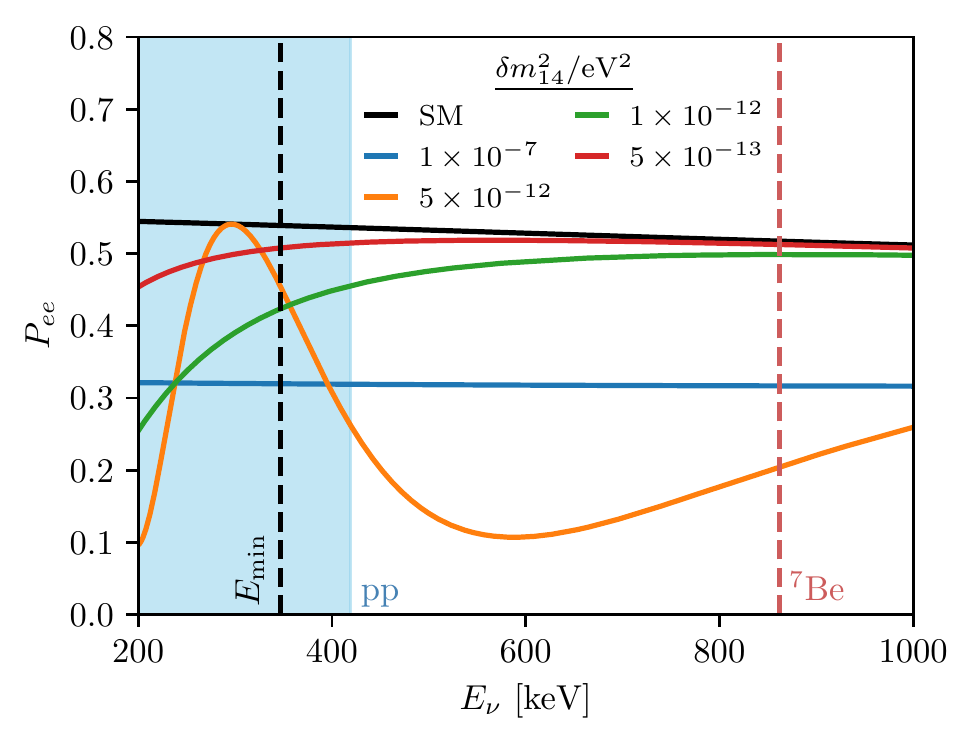}
		\caption{Numerical results for solar electron neutrino survival probability, $P_{ee}$, for the SM scenario (black dashed curve) and for the 1-4 pseudo-Dirac pair scenario with maximal mixing ($\theta_{14}=\pi/4$) and mass splitting $\delta m^2_{14} = 10^{-7}$~eV$^2$ (blue), $5 \times 10^{-12}$~eV$^2$ (orange), $10^{-12}$ eV$^2$ (green), and $5\times 10^{-13}$ eV$^2$ (red). The light blue shaded region corresponds to the energies of $pp$ neutrinos coming from the Sun, and the vertical red dashed line is the monochromatic energy of the high energy $^7$Be neutrino line. The vertical black dashed line is the minimum neutrino energy given a cut in the recoil energy of 200 keV.}
		\label{fig:prob_plot}
	\end{figure}
	%%%%%%%%%%%%%%%%%%%%%%%%%%%%%%%%%%%%%%%%%%%%%%%%%%%%%%%%%%%%%%
	Eq.~\eqref{eq:nu_amp} can be used to calculate the oscillation probabilities at the surface of the Sun. However, for small enough values of the mass splitting, we must consider the vacuum oscillations between the Sun and the Earth. This is because the production region is smaller than the typical oscillation length, and the decoherence length is larger than the distance between the Earth and the Sun, $L_{ES}$. On the other hand, $\Delta m^2_{12}$ and $\Delta m^2_{13}$ are sufficiently large that decoherence between these mass states occurs over distances much smaller than $L_{ES}$. This results in the neutrino mass states decohering into the three mass pairs, which we denote as 1-4, 2-5, and 3-6. We thus have to modify the amplitude at the edge of the Sun, $\mathcal{A}_\odot$, with two mass state vacuum oscillations.
	\begin{gather*}
		\left(\mathcal{A}_E \right)_i = \left(\mathcal{A}_\odot \right)_i \,,\\
		\left(\mathcal{A}_E \right)_{i+3} = \exp\left( -i\frac{\delta m^2 L_{\mathrm{ES}}}{2E_\nu}\right)\left(\mathcal{A}_\odot \right)_{i+3}\,,
	\end{gather*}
	where the index $i = 1,2,3$ denotes the mass state. From the amplitude at Earth, $\mathcal{A}_E$, we can determine the appearance probability of some flavour $\alpha$ to be:
	\be\ba
	P_{e\alpha} = &\left| U_{\alpha 1}\left(\mathcal{A}_E \right)_1 + U_{\alpha 4}\left(\mathcal{A}_E \right)_4\right|^2 \\
	+& \left| U_{\alpha 2}\left(\mathcal{A}_E \right)_2 + U_{\alpha 5}\left(\mathcal{A}_E \right)_5\right|^2 \\
	+ &\left| U_{\alpha 3}\left(\mathcal{A}_E \right)_3 + U_{\alpha 6}\left(\mathcal{A}_E \right)_6\right|^2 .
	\ea\ee
	This modified probability is then the input for the averaged probability used for our analysis. 
	In \figref{fig:prob_plot}, we show the solar electron neutrino survival probability in the 1-4 pair scenario. We observe that for a mass splitting of $\delta m^2_{14} = 10^{-7}$ eV$^2$, the averaging of the probability results in a flat decrease in the survival probability and that the production region is large enough that there is little energy dependence on the probability. As the mass splitting decreases, this no longer is the case and vacuum oscillations dominate, as can be seen for $ \delta m^2_{14} = 5 \times 10^{-12}$ eV$^2$. As the mass splitting decreases, the survival probability slowly approaches the standard oscillations until they are almost indiscernible. This is because the vacuum oscillation length becomes larger than the distance between the Earth and the Sun, leading to a smaller modification to the probability at the surface of the Sun. The oscillation length is proportional to the neutrino energy, and so lower energy neutrinos can probe smaller mass splittings. This will set the limit on the lowest mass splitting that JUNO can probe. 
	
	%%%%%%%%%%%%%%%%%%%%%%%%%%%%%%%%%%%%%%%%%%%%%%%%%%%%%%%%%%%%%%%%%%%%%%%%%%%%%%%%%%%%%%%%%%%%%%%%%%%%%%%%%%%%%%%%%%
	\section{Solar Neutrinos at JUNO}\label{sec:sol} 
	
	JUNO is a multi-purpose neutrino experiment proposed in 2008, with a primary objective to determine the neutrino mass ordering~\cite{JUNO:2015zny}.
	JUNO will constrain this parameter by measuring reactor antineutrinos' survival probability from the Yangjiang and Taishan nuclear power plants (NPPs). 
	The neutrino detector is a liquid scintillator with a 20-kiloton fiducial mass 53 km from the two NPPs. 
	While primarily designed to detect reactor antineutrinos via inverse beta decay, JUNO can also detect solar neutrinos via elastic neutrino electron scattering,
	\begin{equation*}
		\nu_\alpha + e^- \longrightarrow \nu_\alpha + e^-\,,
	\end{equation*}
	where $\alpha$ is the flavour of the incident neutrino. The differential cross-section of this process, to first order in the effective weak interaction, is
	\begin{equation}\label{eq:cross_sec}
		\frac{d\sigma^{(\alpha)}}{d E_r} = \frac{2 G_F^2 m_e}{\pi} \left[ {g^{(\alpha)}_L}^2 + g_R^2 \left(1-\frac{E_r}{E_\nu}\right)^2 - g^{(\alpha)}_L g_R \frac{m_e E_r}{E_\nu^2} \right]\,,
	\end{equation}
	where $G_F$ is the Fermi constant, $E_r$  is the recoil energy  of the outgoing electron, $m_e$ is the electron mass, $E_\nu$ is the energy of the incident neutrino and $g_L^{(\alpha)}$, $g_R$, are the (flavour dependent) coupling constants, which are related to the weak mixing angle $\theta_W$ via
	\begin{table}[t!]
		\centering
		\begin{tabular}{c|c|c|c}
			Source & $\Phi\,$(cm$^{-2}$s$^{-1}$) & $Q\,$(keV) & $A\,$(keV$^{-5}$)\\
			\hline
			$pp$     & $5.98 \times 10^{10}$     & 420      & $1.9232 \times 10^{-13}$\\
			$^7$Be & $4.93 \times 10^9$        & 862, 384 & N/A\\
			\hline
		\end{tabular}
		\caption{Parameters for the differential fluxes of solar neutrino sources used in this work, from \cite{Aalbers:2020}}
		\label{tab:flux_vals}
	\end{table}
	
	\begin{align}
		g_L^{(\alpha)} &= \sin^2 \theta_W - \frac{1}{2} + \delta_{\alpha,e}\,, \\
		g_R &= \sin^2 \theta_W \,.
	\end{align}
	The delta function in flavour space arises from the enhancement of $e^- - \nu_e$ scattering due to the additional charged-current interaction. The differential event rate of measured electrons in the detector can be expressed as \cite{Aalbers:2020}
	\be\label{eq:event_rate}
	\fd{R^{i,a}}{E_r} = N_e \sum_\alpha \fd{\sigma^{(\alpha)}}{E_r} \int dE_\nu P_{e\alpha}(E_\nu)\fd{\phi^{a}}{E_\nu}\,,
	\ee
	where $N_e$ is the number of electrons per kiloton in the target medium, $d\sigma^{(\alpha)}/dE_r$ is the differential cross section for neutrino-electron scattering as shown in Eq.~\eqref{eq:cross_sec}, and $P_{e\alpha}$ is the probability for a neutrino with flavour $\alpha$ arriving at the detector from the Sun. The index $a$ runs over the solar neutrino sources, $pp$ and $^7$Be.
	Their differential spectra $d\phi^a / dE_\nu$ can either be monochromatic in energy as is the case for $^7$Be neutrinos or have a continuous $\beta$ form such as the $pp$ neutrino source:
	\be\label{eq:dflux}
	\fd{\phi}{E_\nu} = \Phi A (x - E_\nu) \left[ (x - E_\nu)^2 - m_e^2 \right]^{\frac{1}{2}} E^2_\nu\,,
	\ee
	where $x = Q + m_e$ with $Q$ being the characteristic energy. The total flux $\Phi$, characteristic energy $Q$, and integral normalisation $A$ are given in table \tabref{tab:flux_vals}. The $^7$Be flux has two monochromatic lines at two different energies, with one at 384 keV making up $10\%$ of the total flux and another at 862 keV contributing the remaining $90\%$ \cite{Lopes:2013,Aalbers:2020}.
	For a neutrino with energy $E_\nu$, the maximum electron recoil energy possible from scattering is given by:
	\begin{equation}
		E_r^{\mathrm{max}} = \frac{2 E^2_\nu}{m_e + 2 E_\nu}\,.
	\end{equation}
	We can equivalently use this relation to find the minimum neutrino energy we must consider when calculating the differential cross section at some recoil energy $E_r$:
	\begin{equation}
		E^\mathrm{min}_\nu = \frac{1}{2} \left( E_r + \sqrt{E^2_r + 2 E_r m_e} \right)\,,
	\end{equation}
	which is the lower integration boundary in \equaref{eq:event_rate} and $Q_a$ is the upper boundary.
	%%%%%%%%%%%%%%%%%%%%%%%%%%%%%%%%%%%%%%%%%%%%%%%%%%%%%%%%%%%%
	\begin{figure}[t!]
		\centering
		\includegraphics[width=0.48\textwidth]{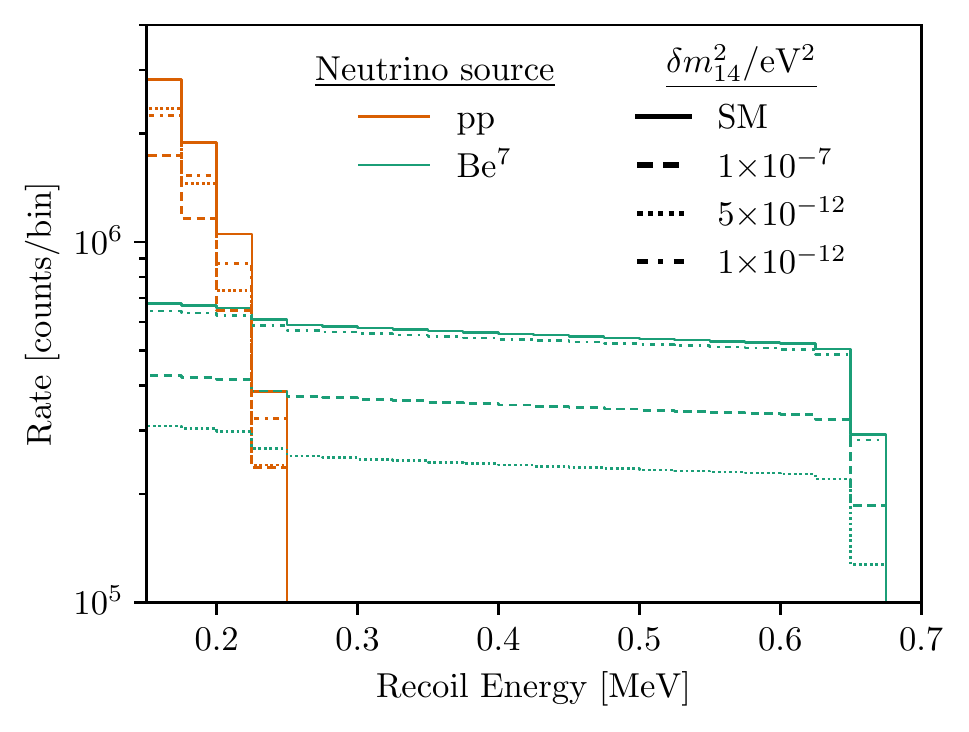}
		\caption{Expected event rates at JUNO for $pp$ (orange) and Be$^7$ (green) solar neutrinos assuming a 6-year exposure time with 20 kiloton fiducial mass. Here we show how a maximal mixing 14 pseudo-Dirac pair affects the event rate that JUNO will measure, for $\delta m^2_{14}$ values of $1\times 10^{-7}$ eV$^2$ (dashed), $5\times 10^{-12}$ eV$^2$ (dotted), $1\times 10^{-12}$ eV$^2$ (dash-dotted), and the SM case (solid). The data is shown with bin widths of 25 keV, as was used in our analysis, to demonstrate how JUNO can put limits on these values.}
		\label{fig:events_plot}
	\end{figure}
	%%%%%%%%%%%%%%%%%%%%%%%%%%%%%%%%%%%%%%%%%%%%%%%%%%%%%%%%%%%%%%
	
	JUNO will be sensitive to both $pp$ and $^7$Be neutrinos and other sources such as $pep$ and CNO neutrinos, though with a lower signal-to-noise ratio. This sensitivity can be used to constrain the parameter space for pseudo-Dirac neutrinos via a solar oscillation analysis which is the objective of this work.
	Nevertheless, the measurement of solar neutrinos in JUNO will depend on the control of backgrounds that affect the low-energy region. 
	Such backgrounds appear due to the resemblance of the neutrino-electron scattering signal to the weak decay of isotopes present in the detector.
	Specifically, a neutrino-electron scattering produces isotropic light with no additional signature, making it indistinguishable from a background one~\cite{JUNO:2021kxb,JUNO:2023zty}.
	In JUNO, the most important background sources are the impurities in the scintillator.
	Other sources can be reduced by choosing a different fiducial volume, for instance ,~\cite {JUNO:2021kxb}.
	The largest background affecting the solar neutrino measurement is the $^{14}$C beta decay process, which completely dominates below 156 keV. If this background is under sufficient control, it can be removed by cutting recoil energies at around 200 keV, which sets a minimum neutrino energy of $\sim$350 keV.
	This allows for the measurement of the high energy part of the $pp$ neutrinos.
	For energies above the $^{14}$C background cut, $^{210}$Bi, $^{85}$Kr, and $^{238}$U will be the main sources of scintillation backgrounds. 
	Since it is still unclear if the $^{14}$C and possible pile-ups would affect recoil energies larger than $\sim 400$ keV, we consider three different situations for the energy threshold in what follows.
	First, an optimistic case where the radiopurity of the scintillator is low enough to have the carbon background and possible pile-ups under control for energies above $200$ keV.
	Second, a more conservative approach where the energy threshold is set to be $450$ keV, similar to the analysis performed by the JUNO collaboration in Ref.~\cite{JUNO:2023zty}.
	We anticipate that the final sensitivity of JUNO will lie between these scenarios, so we have also included a third case for a cut at $250$ keV to demonstrate how the sensitivity may vary.
	
	%%%%%%%%%%%%%%%%%%%%%%%%%%%%%%%%%%%%%%%%%%%%%%%%%%%%%%%%%%%%%%%%%%%%%%%%%%%%%%%%%%%%%%%%%%%%%%%%%%%%%%%%%%%%%%%%%%
	\section{Oscillation  Analysis}\label{sec:osc}
	We aim to quantify JUNO's sensitivity to the pseudo-Dirac neutrino parameter space. To do this, we will calculate the probabilities of the active neutrinos arriving at the detector using the methods discussed in \secref{sec:PDosc}. Given the input parameters, the probabilities will give us the number of events we expect to see at the detector, $N_{\mathrm{theory}}$, as is shown in \figref{fig:events_plot}. 
	The overall effect of pseudo-Dirac oscillations is to reduce the electron neutrino survival probability at Earth since part of the neutrinos would oscillate to invisible sterile states. 
	This is especially clear for the value of $\delta m_{14}^2 = 5\times 10^{-12}~{\rm eV^2}$, where we observe a deficit of $\sim36\%$ with respect to the total expected events in the standard scenario.
	As these give the largest contribution to the cross-section, we can test the pseudo-Dirac scenarios by searching for a decrease in the number of detected neutrino scattering events compared to the SM theoretical expectation.
	The ability of the JUNO experiment to discriminate between the standard and pseudo-Dirac oscillation scenarios is given by the following test statistics,
	\begin{widetext}
		\begin{align}\label{eq:chisq}
			\chi^2 = \sum_{i} \frac{\left( \sum_a \alpha_{a} N^{i,a}_{\mathrm{theory}} + \sum_{b} (\alpha_{b}-1)N^{i,b} - N^i_{\mathrm{bench}} \right)^2}{N^i_{\mathrm{bench}} + \sum_{b} N^i_{b}}
			+ \sum_{a}\left( \frac{\alpha_{a} - 1 }{\sigma_{a}} \right)^2 + \sum_{b} \left( \frac{\alpha_{b} - 1 }{\sigma_{b}} \right)^2\,,
		\end{align}
	\end{widetext}
	which compares the predicted events from the theory and the standard oscillation case. In Eq.~\eqref{eq:chisq}, $N^i$ is the total number of counts in the $i$th recoil energy bin from some source, given a target mass, $M_\mathrm{target}$, and exposure time, $t$. The bin width is taken to be 25 keV, in accordance with the expected energy resolution of JUNO of 3$\%\sqrt{E_r/\mathrm{MeV}}$~\cite{JUNO:2015zny}. The index $b$ runs over the backgrounds for the neutrino detection process, and $N^i_{\mathrm{bench}}$ is the benchmark neutrino event rate expected for the standard oscillation scenario, i.e. $N^i_{\mathrm{bench}} = \sum_a N^{i,a}_{\mathrm{SM}}$. The pull parameters $\alpha_i$ are free parameters that encode the measured events' statistical deviation from the theoretical expectation.
	For this analysis, we fix the standard oscillation parameters at their central values, using the NuFIT 5.2 global fit data \cite{Esteban:2020cvm},  since JUNO is expected to measure independently that the solar parameters $\theta_{12}$ and $\Delta m_{12}^2$ below the percent level using reactor antineutrinos~\cite{JUNO:2015zny}.
	
	The background rates are taken from detector simulations performed by the JUNO collaboration \cite{JUNO:2015zny}. 
	These simulations provide two possible scenarios for the reduction of backgrounds in the detector, the `baseline' case and the `ideal' case. 
	As previously mentioned, the sources of backgrounds for the scintillation signal are from the detector's beta-decay processes of radioactive nuclei. 
	For the ideal case, we consider $^{210}$Bi, $^{85}$Kr, and $^{238}$U as the main backgrounds. These are also very relevant in the baseline case. However, it is also necessary to account for $^{40}$K and $^{232}$Th decay chains.
	
	The pull parameters are given a weighting assuming a Gaussian prior with an error $\sigma$. For the neutrino sources, these errors correspond to the error in the theoretical flux calculations taken from standard solar model (SSM) simulations \cite{Vinyoles:2017}: $\sigma_{\mathrm{pp}} = 0.6\%$ and $\sigma_{\mathrm{^7Be}} = 6\%$. We assume that the background counts can be constrained to a value of $\sigma_{\mathrm{bkg}} = 1\%$, which we believe to be appropriate from the simulations of the backgrounds performed by the JUNO collaboration \cite{JUNO:2015zny}. Minimising the test statistic over the pull parameters will give the projected sensitivity of JUNO to deviations from the standard oscillation scenario. 
	
	%%%%%%%%%%%%%%%%%%%%%%%%%%%%%%%%%%%%%%%%%%%%%%%%%%%%%%%%%%%%%%%%%%%%%%%%%%%%%%%%%%%%%%%%%%%%%%%%%%%%%%%%%%%%%%%%%%%%
	\section{Constraints on the Possible Pseudo-Dirac Nature of Neutrinos from JUNO}\label{sec:res}
	We simulate 6 years of exposure for JUNO, assuming a 20 kiloton fiducial mass. All results are taken for the `ideal' background case. However, the calculations were performed for the 'baseline' case and were found to be similar.
	
	\begin{figure}[t!]
		\centering
		\includegraphics[width=0.48\textwidth]{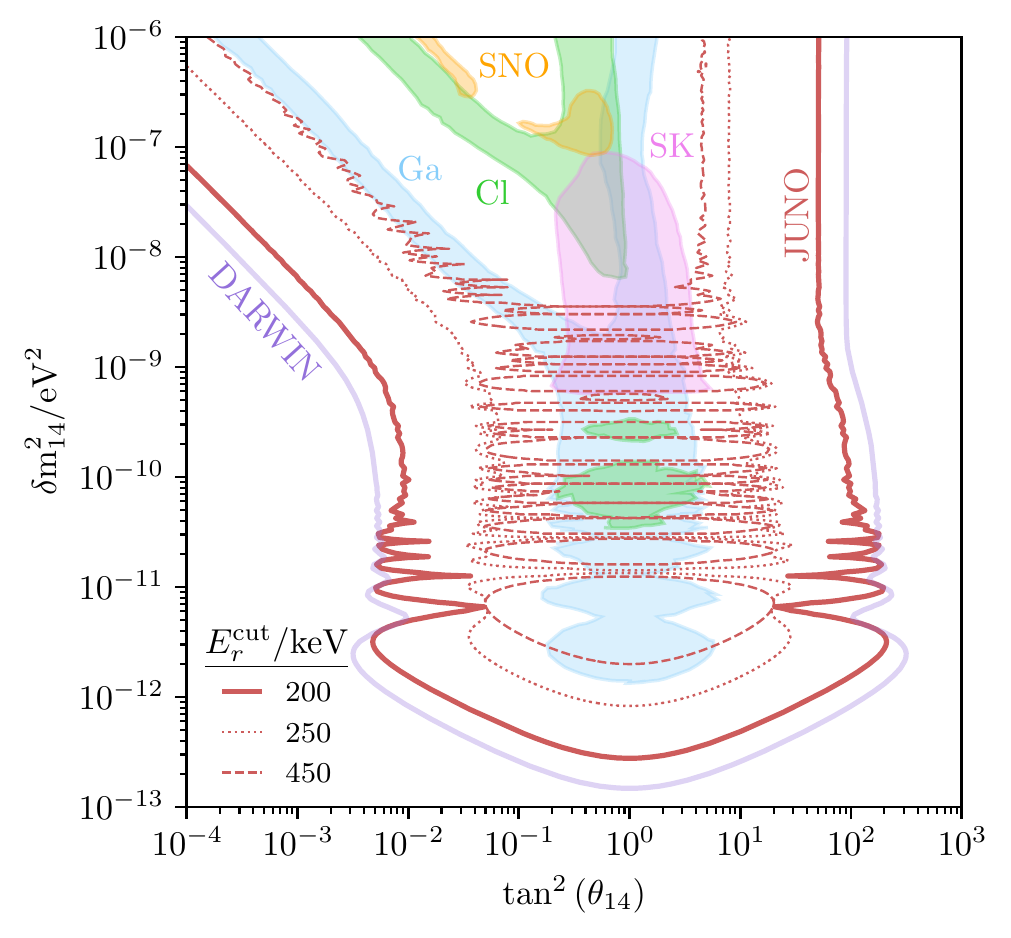}
		\caption{The 2$\sigma$ sensitivity on the parameter space of 1-4 pseudo-Dirac pair oscillations at JUNO (red). These are shown for three values of the recoil energy cut: 200 keV (solid line), 250 keV (dotted line), and 450 keV (dashed line). Also shown are the projected limits that can be set by DARWIN (purple), as well as the current limits from Gadolinium (Ga), Super-Kamiokande (SK), SNO and Cl experiments.}
		\label{fig:JUNO_14_2d}
	\end{figure}
	
	From \figref{fig:JUNO_14_2d} we observe that JUNO can place strong limits on the 1-4 mixing scenario, competing with the capability of the future DARWIN Xenon-based detector that was calculated in \cite{deGouvea:2021ymm}. This is somewhat surprising since the absence of the $^{14}$C background at DARWIN gives it access to much more of the $pp$ neutrino spectrum, which gives stronger bounds on the mixing scenario since it has lower energy and a strongly constrained uncertainty. However, the large fiducial mass of JUNO, 20 kilotons as compared to DARWIN's expected 300 tonnes, means that a large number of $pp$ neutrinos could be detected if the cut at 200 keV in recoil energy is possible. As illustrated in \figref{fig:prob_plot}, the survival probability is much lower for mass splittings $\delta m_{14} ^{2} \gtrsim 10^{-12}$ eV$^2$  the energy range of $pp$ neutrinos than at the $^7$Be energy. This results in a significant difference between the detection rate for $pp$ neutrinos. Furthermore, the theoretical constraints on the $pp$ flux are much tighter than on the $^7$Be flux, resulting in a stronger statistical significance on any deviation from the expected detection rate of $pp$ neutrinos. This allows JUNO to place strong constraints on this scenario.
	\begin{figure}[t!]
		\centering
		\includegraphics[width=0.48\textwidth]{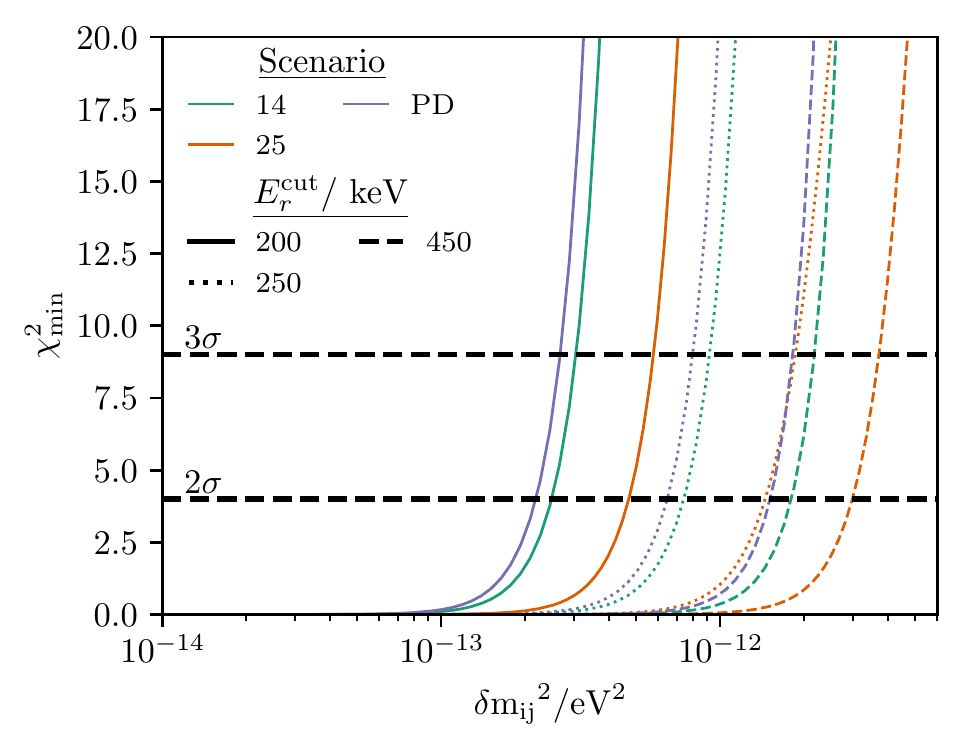}
		\caption{Marginalised $\chi^2$ as a function of the mass splitting $\delta m^2_{ij}$ for various pseudo-Dirac scenarios at JUNO with maximal mixing $\theta_{ij}=\pi/4$. We consider here the 1-4 pair scenario (green), the 2-5 pair scenario (orange), and the full pseudo-Dirac (PD) scenario (purple) with three mass pairs, each with the same mass squared splitting. The sensitivity for each scenario is also displayed for the three recoil cuts of 200 keV (solid lines), 250 keV (dotted lines), and 450 keV (dashed lines)}
		\label{fig:JUNO_1d_all}
	\end{figure}
	
	If JUNO can achieve the intermediate cut (at 250 keV), it will be competitive with the constraints from Borexino. However, if the backgrounds are sufficiently reduced, and the conservative cut is used, then JUNO would only be competitive with Borexino in the $^7$Be neutrino sample, but since  Borexino has measured $pp$ neutrinos using a combined sample, it has marginally greater constraining power than JUNO in this scenario. If, on the other hand, JUNO can achieve the optimistic cut, then it would exceed the sensitivity of Borexino and be competitive with DARWIN.

	We have also considered a maximal mixing angle $\theta = \pi/4$ and determined the constraints JUNO can place on the value of the mass splitting. We computed the sensitivities for the case of 1-4 and 2-5 mixing, as well as for the full pseudo-Dirac scenario assuming that each pair of mass states is split by the same amount, $\delta m^2$. The results of this analysis are shown in \figref{fig:JUNO_1d_all}. In the optimistic scenario we find that JUNO should be capable of excluding a mass splitting above $ \sim 3.1\times 10^{-13} \mathrm{eV}^2$ for the 1-4 scenario and $\sim 6\times 10^{-13} \mathrm{eV}^2$ for the 2-5 scenario with a 3$\sigma$ C.L. The disparity between the two arises because the electron flavour state has a larger component of the 1 - 4 mass state neutrinos than the 2 - 5, so it is more sensitive to the oscillations of the former pair. For the ``full''  pseudo-Dirac case, JUNO would be capable of excluding above $\delta m^2 \gtrsim 2.9\times 10^{-13} \mathrm{eV}^2$, which is lower than either of the two individual cases. This occurs because all of the components of the electron neutrino can oscillate into sterile states, removing the limiting factor of the PMNS mixing and thus increasing the probability of a sterile state being at the detector.
	
	JUNO will also be able to probe the $\delta m^2_{25}$ parameter space, which is important since there have been indications of a preference for a non-zero value of this parameter as in \cite{Ansarifard:2023}. In particular, the preferred value of $\delta m^2 \sim 10^{-11}$ eV$^2$ is testable by JUNO; however, this assumes that the cut at 200 keV in recoil energy is feasible, as for higher cuts there is a dip in sensitivity at around this value. This is due to oscillation effects, where $P_{ee}$ becomes the same as the SM for the higher energy $^7$Be neutrinos. When $pp$ neutrinos are included, these dips are removed since the measured flux is integrated over energy, and the minima are smeared out. The monochromatic nature of the $^7$Be flux could be utilised in a seasonal variation analysis, as was done in \cite{Ansarifard:2023} to search for pseudo-Dirac neutrinos. Due to the large number of these neutrinos that will be detected at JUNO, this could improve on the analysis already done and would be an interesting possibility to explore. We leave this for future work.
	
	%%%%%%%%%%%%%%%%%%%%%%%%%%%%%%%%%%%%%%%%%%%%%%%%%%%%%%%%%%%%%%%%%%%%%%%%%%%%%%%%%%%%%%%%%%%%%%%%%%%%%%%%%%%%%%%%%%%%
	\section{Summary and Conclusions}\label{sec:conc}
	
	If super-light sterile neutrinos exist, they could significantly change the survival probability of solar neutrinos. We consider two scenarios:  $3+1$, where one additional light state has been added to the neutrino spectrum, and the  $3+3$ scenario, where there are three new light states. To determine the sensitivity of JUNO to these pseudo-Dirac neutrino scenarios,  we use analytic expressions for the electron neutrino survival probability for $pp$-neutrinos and perform a numerical diagonalisation of the Hamiltonian in matter for the higher energy $^7$Be neutrinos to determine the neutrinos detectable by JUNO for a given point in the theory parameter space. Additional light states generally decrease the electron neutrino survival probability, as shown in \figref{fig:events_plot}.

	Using publicly available backgrounds from the JUNO collaboration, we evaluate the statistical significance of the new physics scenario for a six-year run time with 20 kiloton fiducial mass. The sensitivity of JUNO to such new physics scenarios varies on the energy threshold of the recoiling electron produced in neutrino-electron elastic scattering. This minimal threshold will ultimately be determined by how well the backgrounds from radio impurities can be controlled. We investigate three scenarios: In the first ``optimistic'' scenario, the radiopurity is low enough to control the carbon background and pile-ups for energies above 200 keV. In the second ``conservative'' scenario,   the energy threshold is set to 450 keV, and finally, we consider an ``intermediate'' scenario where we place the cuts at  250 keV. 
	In the conservative scenario, the sensitivity of JUNO to the pseudo-Dirac parameter space is competitive with limits from Borexino using its $^7$Be neutrino sample \cite{Chen:2022bxn}.   However, if JUNO can control its backgrounds to the level that the cut can be placed at 200 (250) keV, then sensitivity can be improved by an order of magnitude (half an order of magnitude). In the $3+1$ scenario, JUNO's sensitivity can be competitive with the proposed experiment DARWIN \cite{DARWIN:2016hyl} if the optimistic scenario is achieved, as shown in \figref{fig:JUNO_14_2d}. We have also quantified the sensitivity of JUNO to the 3 + 3 scenario (see \figref{fig:JUNO_1d_all}) and found that $\delta m^2 \gtrsim 2.9\times 10^{-13}\,\rm{eV}^2$ can be excluded at the 3$\sigma$ level in the optimistic scenario. Overall, our results suggest that JUNO has great potential to explore the pseudo-Dirac parameter space and shed light on the existence of super-light sterile neutrinos.
	
	%%%%%%%%%%%%%%%%%%%%%%%%%%%%%%%%%%%%%%%%%%%%%%%%%%%%%%%%%%%%%%%%%%%%%%%%%%%%%%%%%%%%%%%%%%%%%%%%%%%%%%%%%%%%%%%%%%%%
	\section*{Acknowledgements}
	We thank Xianguo Lu and Liangjian Wen for their advice on our JUNO-related questions. 
	In addition, we are grateful to Ivan Martinez-Soler for insightful discussions on this work. 
	This work has made use of the Hamilton HPC Service of Durham University.
	
	\bibliographystyle{apsrev4-1}
	\bibliography{ref.bib}

\begin{thebibliography}{49}%
\makeatletter
\providecommand \@ifxundefined [1]{%
 \@ifx{#1\undefined}
}%
\providecommand \@ifnum [1]{%
 \ifnum #1\expandafter \@firstoftwo
 \else \expandafter \@secondoftwo
 \fi
}%
\providecommand \@ifx [1]{%
 \ifx #1\expandafter \@firstoftwo
 \else \expandafter \@secondoftwo
 \fi
}%
\providecommand \natexlab [1]{#1}%
\providecommand \enquote  [1]{``#1''}%
\providecommand \bibnamefont  [1]{#1}%
\providecommand \bibfnamefont [1]{#1}%
\providecommand \citenamefont [1]{#1}%
\providecommand \href@noop [0]{\@secondoftwo}%
\providecommand \href [0]{\begingroup \@sanitize@url \@href}%
\providecommand \@href[1]{\@@startlink{#1}\@@href}%
\providecommand \@@href[1]{\endgroup#1\@@endlink}%
\providecommand \@sanitize@url [0]{\catcode `\\12\catcode `\$12\catcode
  `\&12\catcode `\#12\catcode `\^12\catcode `\_12\catcode `\%12\relax}%
\providecommand \@@startlink[1]{}%
\providecommand \@@endlink[0]{}%
\providecommand \url  [0]{\begingroup\@sanitize@url \@url }%
\providecommand \@url [1]{\endgroup\@href {#1}{\urlprefix }}%
\providecommand \urlprefix  [0]{URL }%
\providecommand \Eprint [0]{\href }%
\providecommand \doibase [0]{http://dx.doi.org/}%
\providecommand \selectlanguage [0]{\@gobble}%
\providecommand \bibinfo  [0]{\@secondoftwo}%
\providecommand \bibfield  [0]{\@secondoftwo}%
\providecommand \translation [1]{[#1]}%
\providecommand \BibitemOpen [0]{}%
\providecommand \bibitemStop [0]{}%
\providecommand \bibitemNoStop [0]{.\EOS\space}%
\providecommand \EOS [0]{\spacefactor3000\relax}%
\providecommand \BibitemShut  [1]{\csname bibitem#1\endcsname}%
\let\auto@bib@innerbib\@empty
%</preamble>
\bibitem [{\citenamefont {Fukuda}\ \emph {et~al.}(1998)\citenamefont {Fukuda}
  \emph {et~al.}}]{Super-Kamiokande:1998kpq}%
  \BibitemOpen
  \bibfield  {author} {\bibinfo {author} {\bibfnamefont {Y.}~\bibnamefont
  {Fukuda}} \emph {et~al.} (\bibinfo {collaboration} {Super-Kamiokande}),\
  }\href {\doibase 10.1103/PhysRevLett.81.1562} {\bibfield  {journal} {\bibinfo
   {journal} {Phys. Rev. Lett.}\ }\textbf {\bibinfo {volume} {81}},\ \bibinfo
  {pages} {1562} (\bibinfo {year} {1998})},\ \Eprint
  {http://arxiv.org/abs/hep-ex/9807003} {arXiv:hep-ex/9807003} \BibitemShut
  {NoStop}%
\bibitem [{\citenamefont {Ahmad}\ \emph {et~al.}(2002)\citenamefont {Ahmad}
  \emph {et~al.}}]{SNO:2002tuh}%
  \BibitemOpen
  \bibfield  {author} {\bibinfo {author} {\bibfnamefont {Q.~R.}\ \bibnamefont
  {Ahmad}} \emph {et~al.} (\bibinfo {collaboration} {SNO}),\ }\href {\doibase
  10.1103/PhysRevLett.89.011301} {\bibfield  {journal} {\bibinfo  {journal}
  {Phys. Rev. Lett.}\ }\textbf {\bibinfo {volume} {89}},\ \bibinfo {pages}
  {011301} (\bibinfo {year} {2002})},\ \Eprint
  {http://arxiv.org/abs/nucl-ex/0204008} {arXiv:nucl-ex/0204008} \BibitemShut
  {NoStop}%
\bibitem [{\citenamefont {Mohapatra}\ and\ \citenamefont
  {Senjanovic}(1980)}]{Mohapatra:1979ia}%
  \BibitemOpen
  \bibfield  {author} {\bibinfo {author} {\bibfnamefont {R.~N.}\ \bibnamefont
  {Mohapatra}}\ and\ \bibinfo {author} {\bibfnamefont {G.}~\bibnamefont
  {Senjanovic}},\ }\href {\doibase 10.1103/PhysRevLett.44.912} {\bibfield
  {journal} {\bibinfo  {journal} {Phys. Rev. Lett.}\ }\textbf {\bibinfo
  {volume} {44}},\ \bibinfo {pages} {912} (\bibinfo {year} {1980})}\BibitemShut
  {NoStop}%
\bibitem [{\citenamefont {Gell-Mann}\ \emph {et~al.}(1979)\citenamefont
  {Gell-Mann}, \citenamefont {Ramond},\ and\ \citenamefont
  {Slansky}}]{Gell-Mann:1979vob}%
  \BibitemOpen
  \bibfield  {author} {\bibinfo {author} {\bibfnamefont {M.}~\bibnamefont
  {Gell-Mann}}, \bibinfo {author} {\bibfnamefont {P.}~\bibnamefont {Ramond}}, \
  and\ \bibinfo {author} {\bibfnamefont {R.}~\bibnamefont {Slansky}},\
  }\href@noop {} {\bibfield  {journal} {\bibinfo  {journal} {Conf. Proc. C}\
  }\textbf {\bibinfo {volume} {790927}},\ \bibinfo {pages} {315} (\bibinfo
  {year} {1979})},\ \Eprint {http://arxiv.org/abs/1306.4669} {arXiv:1306.4669
  [hep-th]} \BibitemShut {NoStop}%
\bibitem [{\citenamefont {Yanagida}(1979)}]{Yanagida:1979as}%
  \BibitemOpen
  \bibfield  {author} {\bibinfo {author} {\bibfnamefont {T.}~\bibnamefont
  {Yanagida}},\ }\href@noop {} {\bibfield  {journal} {\bibinfo  {journal}
  {Conf. Proc. C}\ }\textbf {\bibinfo {volume} {7902131}},\ \bibinfo {pages}
  {95} (\bibinfo {year} {1979})}\BibitemShut {NoStop}%
\bibitem [{\citenamefont {Minkowski}(1977)}]{Minkowski:1977sc}%
  \BibitemOpen
  \bibfield  {author} {\bibinfo {author} {\bibfnamefont {P.}~\bibnamefont
  {Minkowski}},\ }\href {\doibase 10.1016/0370-2693(77)90435-X} {\bibfield
  {journal} {\bibinfo  {journal} {Phys. Lett. B}\ }\textbf {\bibinfo {volume}
  {67}},\ \bibinfo {pages} {421} (\bibinfo {year} {1977})}\BibitemShut
  {NoStop}%
\bibitem [{\citenamefont {Mohapatra}\ and\ \citenamefont
  {Senjanovic}(1981)}]{Mohapatra:1980yp}%
  \BibitemOpen
  \bibfield  {author} {\bibinfo {author} {\bibfnamefont {R.~N.}\ \bibnamefont
  {Mohapatra}}\ and\ \bibinfo {author} {\bibfnamefont {G.}~\bibnamefont
  {Senjanovic}},\ }\href {\doibase 10.1103/PhysRevD.23.165} {\bibfield
  {journal} {\bibinfo  {journal} {Phys. Rev. D}\ }\textbf {\bibinfo {volume}
  {23}},\ \bibinfo {pages} {165} (\bibinfo {year} {1981})}\BibitemShut
  {NoStop}%
\bibitem [{\citenamefont {Magg}\ and\ \citenamefont
  {Wetterich}(1980)}]{Magg:1980ut}%
  \BibitemOpen
  \bibfield  {author} {\bibinfo {author} {\bibfnamefont {M.}~\bibnamefont
  {Magg}}\ and\ \bibinfo {author} {\bibfnamefont {C.}~\bibnamefont
  {Wetterich}},\ }\href {\doibase 10.1016/0370-2693(80)90825-4} {\bibfield
  {journal} {\bibinfo  {journal} {Phys. Lett. B}\ }\textbf {\bibinfo {volume}
  {94}},\ \bibinfo {pages} {61} (\bibinfo {year} {1980})}\BibitemShut {NoStop}%
\bibitem [{\citenamefont {Lazarides}\ \emph {et~al.}(1981)\citenamefont
  {Lazarides}, \citenamefont {Shafi},\ and\ \citenamefont
  {Wetterich}}]{Lazarides:1980nt}%
  \BibitemOpen
  \bibfield  {author} {\bibinfo {author} {\bibfnamefont {G.}~\bibnamefont
  {Lazarides}}, \bibinfo {author} {\bibfnamefont {Q.}~\bibnamefont {Shafi}}, \
  and\ \bibinfo {author} {\bibfnamefont {C.}~\bibnamefont {Wetterich}},\ }\href
  {\doibase 10.1016/0550-3213(81)90354-0} {\bibfield  {journal} {\bibinfo
  {journal} {Nucl. Phys. B}\ }\textbf {\bibinfo {volume} {181}},\ \bibinfo
  {pages} {287} (\bibinfo {year} {1981})}\BibitemShut {NoStop}%
\bibitem [{\citenamefont {Wetterich}(1981)}]{Wetterich:1981bx}%
  \BibitemOpen
  \bibfield  {author} {\bibinfo {author} {\bibfnamefont {C.}~\bibnamefont
  {Wetterich}},\ }\href {\doibase 10.1016/0550-3213(81)90279-0} {\bibfield
  {journal} {\bibinfo  {journal} {Nucl. Phys. B}\ }\textbf {\bibinfo {volume}
  {187}},\ \bibinfo {pages} {343} (\bibinfo {year} {1981})}\BibitemShut
  {NoStop}%
\bibitem [{\citenamefont {Foot}\ \emph {et~al.}(1989)\citenamefont {Foot},
  \citenamefont {Lew}, \citenamefont {He},\ and\ \citenamefont
  {Joshi}}]{Foot:1988aq}%
  \BibitemOpen
  \bibfield  {author} {\bibinfo {author} {\bibfnamefont {R.}~\bibnamefont
  {Foot}}, \bibinfo {author} {\bibfnamefont {H.}~\bibnamefont {Lew}}, \bibinfo
  {author} {\bibfnamefont {X.~G.}\ \bibnamefont {He}}, \ and\ \bibinfo {author}
  {\bibfnamefont {G.~C.}\ \bibnamefont {Joshi}},\ }\href {\doibase
  10.1007/BF01415558} {\bibfield  {journal} {\bibinfo  {journal} {Z. Phys. C}\
  }\textbf {\bibinfo {volume} {44}},\ \bibinfo {pages} {441} (\bibinfo {year}
  {1989})}\BibitemShut {NoStop}%
\bibitem [{\citenamefont {Ma}(1998)}]{Ma:1998dn}%
  \BibitemOpen
  \bibfield  {author} {\bibinfo {author} {\bibfnamefont {E.}~\bibnamefont
  {Ma}},\ }\href {\doibase 10.1103/PhysRevLett.81.1171} {\bibfield  {journal}
  {\bibinfo  {journal} {Phys. Rev. Lett.}\ }\textbf {\bibinfo {volume} {81}},\
  \bibinfo {pages} {1171} (\bibinfo {year} {1998})},\ \Eprint
  {http://arxiv.org/abs/hep-ph/9805219} {arXiv:hep-ph/9805219} \BibitemShut
  {NoStop}%
\bibitem [{\citenamefont {Wolfenstein}(1981)}]{Wolfenstein:1981kw}%
  \BibitemOpen
  \bibfield  {author} {\bibinfo {author} {\bibfnamefont {L.}~\bibnamefont
  {Wolfenstein}},\ }\href {\doibase 10.1016/0550-3213(81)90096-1} {\bibfield
  {journal} {\bibinfo  {journal} {Nucl. Phys. B}\ }\textbf {\bibinfo {volume}
  {186}},\ \bibinfo {pages} {147} (\bibinfo {year} {1981})}\BibitemShut
  {NoStop}%
\bibitem [{\citenamefont {Petcov}(1982)}]{Petcov:1982ya}%
  \BibitemOpen
  \bibfield  {author} {\bibinfo {author} {\bibfnamefont {S.~T.}\ \bibnamefont
  {Petcov}},\ }\href {\doibase 10.1016/0370-2693(82)91246-1} {\bibfield
  {journal} {\bibinfo  {journal} {Phys. Lett. B}\ }\textbf {\bibinfo {volume}
  {110}},\ \bibinfo {pages} {245} (\bibinfo {year} {1982})}\BibitemShut
  {NoStop}%
\bibitem [{\citenamefont {Bilenky}\ and\ \citenamefont
  {Pontecorvo}(1983)}]{Bilenky:1983wt}%
  \BibitemOpen
  \bibfield  {author} {\bibinfo {author} {\bibfnamefont {S.~M.}\ \bibnamefont
  {Bilenky}}\ and\ \bibinfo {author} {\bibfnamefont {B.}~\bibnamefont
  {Pontecorvo}},\ }\href@noop {} {\bibfield  {journal} {\bibinfo  {journal}
  {Sov. J. Nucl. Phys.}\ }\textbf {\bibinfo {volume} {38}},\ \bibinfo {pages}
  {248} (\bibinfo {year} {1983})}\BibitemShut {NoStop}%
\bibitem [{\citenamefont {Foot}\ and\ \citenamefont
  {Volkas}(1995)}]{Foot:1995pa}%
  \BibitemOpen
  \bibfield  {author} {\bibinfo {author} {\bibfnamefont {R.}~\bibnamefont
  {Foot}}\ and\ \bibinfo {author} {\bibfnamefont {R.~R.}\ \bibnamefont
  {Volkas}},\ }\href {\doibase 10.1103/PhysRevD.52.6595} {\bibfield  {journal}
  {\bibinfo  {journal} {Phys. Rev. D}\ }\textbf {\bibinfo {volume} {52}},\
  \bibinfo {pages} {6595} (\bibinfo {year} {1995})},\ \Eprint
  {http://arxiv.org/abs/hep-ph/9505359} {arXiv:hep-ph/9505359} \BibitemShut
  {NoStop}%
\bibitem [{\citenamefont {Chang}\ and\ \citenamefont
  {Kong}(2000)}]{Chang:1999pb}%
  \BibitemOpen
  \bibfield  {author} {\bibinfo {author} {\bibfnamefont {D.}~\bibnamefont
  {Chang}}\ and\ \bibinfo {author} {\bibfnamefont {O.~C.~W.}\ \bibnamefont
  {Kong}},\ }\href {\doibase 10.1016/S0370-2693(00)00228-8} {\bibfield
  {journal} {\bibinfo  {journal} {Phys. Lett. B}\ }\textbf {\bibinfo {volume}
  {477}},\ \bibinfo {pages} {416} (\bibinfo {year} {2000})},\ \Eprint
  {http://arxiv.org/abs/hep-ph/9912268} {arXiv:hep-ph/9912268} \BibitemShut
  {NoStop}%
\bibitem [{\citenamefont {Carloni}\ \emph {et~al.}(2022)\citenamefont
  {Carloni}, \citenamefont {Martinez-Soler}, \citenamefont {Arguelles},
  \citenamefont {Babu},\ and\ \citenamefont {Dev}}]{Carloni:2022cqz}%
  \BibitemOpen
  \bibfield  {author} {\bibinfo {author} {\bibfnamefont {K.}~\bibnamefont
  {Carloni}}, \bibinfo {author} {\bibfnamefont {I.}~\bibnamefont
  {Martinez-Soler}}, \bibinfo {author} {\bibfnamefont {C.~A.}\ \bibnamefont
  {Arguelles}}, \bibinfo {author} {\bibfnamefont {K.~S.}\ \bibnamefont {Babu}},
  \ and\ \bibinfo {author} {\bibfnamefont {P.~S.~B.}\ \bibnamefont {Dev}},\
  }\href@noop {} {\  (\bibinfo {year} {2022})},\ \Eprint
  {http://arxiv.org/abs/2212.00737} {arXiv:2212.00737 [astro-ph.HE]}
  \BibitemShut {NoStop}%
\bibitem [{\citenamefont {Mohapatra}(1987)}]{Mohapatra:1987hh}%
  \BibitemOpen
  \bibfield  {author} {\bibinfo {author} {\bibfnamefont {R.~N.}\ \bibnamefont
  {Mohapatra}},\ }\href {\doibase 10.1016/0370-2693(87)90161-4} {\bibfield
  {journal} {\bibinfo  {journal} {Phys. Lett. B}\ }\textbf {\bibinfo {volume}
  {198}},\ \bibinfo {pages} {69} (\bibinfo {year} {1987})}\BibitemShut
  {NoStop}%
\bibitem [{\citenamefont {Babu}\ and\ \citenamefont {He}(1989)}]{Babu:1988yq}%
  \BibitemOpen
  \bibfield  {author} {\bibinfo {author} {\bibfnamefont {K.~S.}\ \bibnamefont
  {Babu}}\ and\ \bibinfo {author} {\bibfnamefont {X.~G.}\ \bibnamefont {He}},\
  }\href {\doibase 10.1142/S0217732389000095} {\bibfield  {journal} {\bibinfo
  {journal} {Mod. Phys. Lett. A}\ }\textbf {\bibinfo {volume} {4}},\ \bibinfo
  {pages} {61} (\bibinfo {year} {1989})}\BibitemShut {NoStop}%
\bibitem [{\citenamefont {Farzan}\ and\ \citenamefont
  {Ma}(2012)}]{Farzan:2012sa}%
  \BibitemOpen
  \bibfield  {author} {\bibinfo {author} {\bibfnamefont {Y.}~\bibnamefont
  {Farzan}}\ and\ \bibinfo {author} {\bibfnamefont {E.}~\bibnamefont {Ma}},\
  }\href {\doibase 10.1103/PhysRevD.86.033007} {\bibfield  {journal} {\bibinfo
  {journal} {Phys. Rev. D}\ }\textbf {\bibinfo {volume} {86}},\ \bibinfo
  {pages} {033007} (\bibinfo {year} {2012})},\ \Eprint
  {http://arxiv.org/abs/1204.4890} {arXiv:1204.4890 [hep-ph]} \BibitemShut
  {NoStop}%
\bibitem [{\citenamefont {Ma}\ and\ \citenamefont {Popov}(2017)}]{Ma:2016mwh}%
  \BibitemOpen
  \bibfield  {author} {\bibinfo {author} {\bibfnamefont {E.}~\bibnamefont
  {Ma}}\ and\ \bibinfo {author} {\bibfnamefont {O.}~\bibnamefont {Popov}},\
  }\href {\doibase 10.1016/j.physletb.2016.11.027} {\bibfield  {journal}
  {\bibinfo  {journal} {Phys. Lett. B}\ }\textbf {\bibinfo {volume} {764}},\
  \bibinfo {pages} {142} (\bibinfo {year} {2017})},\ \Eprint
  {http://arxiv.org/abs/1609.02538} {arXiv:1609.02538 [hep-ph]} \BibitemShut
  {NoStop}%
\bibitem [{\citenamefont {Saad}(2019)}]{Saad:2019bqf}%
  \BibitemOpen
  \bibfield  {author} {\bibinfo {author} {\bibfnamefont {S.}~\bibnamefont
  {Saad}},\ }\href {\doibase 10.1016/j.nuclphysb.2019.114636} {\bibfield
  {journal} {\bibinfo  {journal} {Nucl. Phys. B}\ }\textbf {\bibinfo {volume}
  {943}},\ \bibinfo {pages} {114636} (\bibinfo {year} {2019})},\ \Eprint
  {http://arxiv.org/abs/1902.07259} {arXiv:1902.07259 [hep-ph]} \BibitemShut
  {NoStop}%
\bibitem [{\citenamefont {Jana}\ \emph {et~al.}(2019)\citenamefont {Jana},
  \citenamefont {Vishnu},\ and\ \citenamefont {Saad}}]{Jana:2019mez}%
  \BibitemOpen
  \bibfield  {author} {\bibinfo {author} {\bibfnamefont {S.}~\bibnamefont
  {Jana}}, \bibinfo {author} {\bibfnamefont {P.~K.}\ \bibnamefont {Vishnu}}, \
  and\ \bibinfo {author} {\bibfnamefont {S.}~\bibnamefont {Saad}},\ }\href
  {\doibase 10.1140/epjc/s10052-019-7441-9} {\bibfield  {journal} {\bibinfo
  {journal} {Eur. Phys. J. C}\ }\textbf {\bibinfo {volume} {79}},\ \bibinfo
  {pages} {916} (\bibinfo {year} {2019})},\ \Eprint
  {http://arxiv.org/abs/1904.07407} {arXiv:1904.07407 [hep-ph]} \BibitemShut
  {NoStop}%
\bibitem [{\citenamefont {Babu}\ \emph {et~al.}(2022)\citenamefont {Babu},
  \citenamefont {He}, \citenamefont {Su},\ and\ \citenamefont
  {Thapa}}]{Babu:2022ikf}%
  \BibitemOpen
  \bibfield  {author} {\bibinfo {author} {\bibfnamefont {K.~S.}\ \bibnamefont
  {Babu}}, \bibinfo {author} {\bibfnamefont {X.-G.}\ \bibnamefont {He}},
  \bibinfo {author} {\bibfnamefont {M.}~\bibnamefont {Su}}, \ and\ \bibinfo
  {author} {\bibfnamefont {A.}~\bibnamefont {Thapa}},\ }\href {\doibase
  10.1007/JHEP08(2022)140} {\bibfield  {journal} {\bibinfo  {journal} {JHEP}\
  }\textbf {\bibinfo {volume} {08}},\ \bibinfo {pages} {140} (\bibinfo {year}
  {2022})},\ \Eprint {http://arxiv.org/abs/2205.09127} {arXiv:2205.09127
  [hep-ph]} \BibitemShut {NoStop}%
\bibitem [{\citenamefont {Davidson}\ and\ \citenamefont
  {Wali}(1987)}]{Davidson:1987mh}%
  \BibitemOpen
  \bibfield  {author} {\bibinfo {author} {\bibfnamefont {A.}~\bibnamefont
  {Davidson}}\ and\ \bibinfo {author} {\bibfnamefont {K.~C.}\ \bibnamefont
  {Wali}},\ }\href {\doibase 10.1103/PhysRevLett.59.393} {\bibfield  {journal}
  {\bibinfo  {journal} {Phys. Rev. Lett.}\ }\textbf {\bibinfo {volume} {59}},\
  \bibinfo {pages} {393} (\bibinfo {year} {1987})}\BibitemShut {NoStop}%
\bibitem [{\citenamefont {Balaji}\ \emph {et~al.}(2002)\citenamefont {Balaji},
  \citenamefont {Kalliomaki},\ and\ \citenamefont {Maalampi}}]{Balaji:2001fi}%
  \BibitemOpen
  \bibfield  {author} {\bibinfo {author} {\bibfnamefont {K.~R.~S.}\
  \bibnamefont {Balaji}}, \bibinfo {author} {\bibfnamefont {A.}~\bibnamefont
  {Kalliomaki}}, \ and\ \bibinfo {author} {\bibfnamefont {J.}~\bibnamefont
  {Maalampi}},\ }\href {\doibase 10.1016/S0370-2693(01)01356-9} {\bibfield
  {journal} {\bibinfo  {journal} {Phys. Lett. B}\ }\textbf {\bibinfo {volume}
  {524}},\ \bibinfo {pages} {153} (\bibinfo {year} {2002})},\ \Eprint
  {http://arxiv.org/abs/hep-ph/0110314} {arXiv:hep-ph/0110314} \BibitemShut
  {NoStop}%
\bibitem [{\citenamefont {Borah}\ and\ \citenamefont
  {Dasgupta}(2017)}]{Borah:2017leo}%
  \BibitemOpen
  \bibfield  {author} {\bibinfo {author} {\bibfnamefont {D.}~\bibnamefont
  {Borah}}\ and\ \bibinfo {author} {\bibfnamefont {A.}~\bibnamefont
  {Dasgupta}},\ }\href {\doibase 10.1088/1475-7516/2017/06/003} {\bibfield
  {journal} {\bibinfo  {journal} {JCAP}\ }\textbf {\bibinfo {volume} {06}},\
  \bibinfo {pages} {003} (\bibinfo {year} {2017})},\ \Eprint
  {http://arxiv.org/abs/1702.02877} {arXiv:1702.02877 [hep-ph]} \BibitemShut
  {NoStop}%
\bibitem [{\citenamefont {Chen}\ \emph {et~al.}(2022)\citenamefont {Chen},
  \citenamefont {Liao}, \citenamefont {Ling},\ and\ \citenamefont
  {Yue}}]{Chen:2022bxn}%
  \BibitemOpen
  \bibfield  {author} {\bibinfo {author} {\bibfnamefont {Z.}~\bibnamefont
  {Chen}}, \bibinfo {author} {\bibfnamefont {J.}~\bibnamefont {Liao}}, \bibinfo
  {author} {\bibfnamefont {J.}~\bibnamefont {Ling}}, \ and\ \bibinfo {author}
  {\bibfnamefont {B.}~\bibnamefont {Yue}},\ }\href {\doibase
  10.1007/JHEP09(2022)004} {\bibfield  {journal} {\bibinfo  {journal} {J. High
  Energ. Phys.}\ } (\bibinfo {year} {2022}),\
  10.1007/JHEP09(2022)004}\BibitemShut {NoStop}%
\bibitem [{\citenamefont {Ansarifard}\ and\ \citenamefont
  {Farzan}(2023)}]{Ansarifard:2023}%
  \BibitemOpen
  \bibfield  {author} {\bibinfo {author} {\bibfnamefont {S.}~\bibnamefont
  {Ansarifard}}\ and\ \bibinfo {author} {\bibfnamefont {Y.}~\bibnamefont
  {Farzan}},\ }\href@noop {} {\  (\bibinfo {year} {2023})},\ \Eprint
  {http://arxiv.org/abs/2211.09105} {arXiv:2211.09105 [hep-ph]} \BibitemShut
  {NoStop}%
\bibitem [{\citenamefont {de~Gouv\^ea}\ \emph {et~al.}(2022)\citenamefont
  {de~Gouv\^ea}, \citenamefont {McGinness}, \citenamefont {Martinez-Soler},\
  and\ \citenamefont {Perez-Gonzalez}}]{deGouvea:2021ymm}%
  \BibitemOpen
  \bibfield  {author} {\bibinfo {author} {\bibfnamefont {A.}~\bibnamefont
  {de~Gouv\^ea}}, \bibinfo {author} {\bibfnamefont {E.}~\bibnamefont
  {McGinness}}, \bibinfo {author} {\bibfnamefont {I.}~\bibnamefont
  {Martinez-Soler}}, \ and\ \bibinfo {author} {\bibfnamefont {Y.~F.}\
  \bibnamefont {Perez-Gonzalez}},\ }\href {\doibase
  10.1103/PhysRevD.106.096017} {\bibfield  {journal} {\bibinfo  {journal}
  {Phys. Rev. D}\ }\textbf {\bibinfo {volume} {106}},\ \bibinfo {pages}
  {096017} (\bibinfo {year} {2022})},\ \Eprint
  {http://arxiv.org/abs/2111.02421} {arXiv:2111.02421 [hep-ph]} \BibitemShut
  {NoStop}%
\bibitem [{\citenamefont {Beacom}\ \emph {et~al.}(2004)\citenamefont {Beacom},
  \citenamefont {Bell}, \citenamefont {Hooper}, \citenamefont {Learned},
  \citenamefont {Pakvasa},\ and\ \citenamefont {Weiler}}]{Beacom:2003eu}%
  \BibitemOpen
  \bibfield  {author} {\bibinfo {author} {\bibfnamefont {J.~F.}\ \bibnamefont
  {Beacom}}, \bibinfo {author} {\bibfnamefont {N.~F.}\ \bibnamefont {Bell}},
  \bibinfo {author} {\bibfnamefont {D.}~\bibnamefont {Hooper}}, \bibinfo
  {author} {\bibfnamefont {J.~G.}\ \bibnamefont {Learned}}, \bibinfo {author}
  {\bibfnamefont {S.}~\bibnamefont {Pakvasa}}, \ and\ \bibinfo {author}
  {\bibfnamefont {T.~J.}\ \bibnamefont {Weiler}},\ }\href {\doibase
  10.1103/PhysRevLett.92.011101} {\bibfield  {journal} {\bibinfo  {journal}
  {Phys. Rev. Lett.}\ }\textbf {\bibinfo {volume} {92}},\ \bibinfo {pages}
  {011101} (\bibinfo {year} {2004})},\ \Eprint
  {http://arxiv.org/abs/hep-ph/0307151} {arXiv:hep-ph/0307151} \BibitemShut
  {NoStop}%
\bibitem [{\citenamefont {Martinez-Soler}\ \emph {et~al.}(2022)\citenamefont
  {Martinez-Soler}, \citenamefont {Perez-Gonzalez},\ and\ \citenamefont
  {Sen}}]{Martinez-Soler:2021unz}%
  \BibitemOpen
  \bibfield  {author} {\bibinfo {author} {\bibfnamefont {I.}~\bibnamefont
  {Martinez-Soler}}, \bibinfo {author} {\bibfnamefont {Y.~F.}\ \bibnamefont
  {Perez-Gonzalez}}, \ and\ \bibinfo {author} {\bibfnamefont {M.}~\bibnamefont
  {Sen}},\ }\href {\doibase 10.1103/PhysRevD.105.095019} {\bibfield  {journal}
  {\bibinfo  {journal} {Phys. Rev. D}\ }\textbf {\bibinfo {volume} {105}},\
  \bibinfo {pages} {095019} (\bibinfo {year} {2022})},\ \Eprint
  {http://arxiv.org/abs/2105.12736} {arXiv:2105.12736 [hep-ph]} \BibitemShut
  {NoStop}%
\bibitem [{\citenamefont {De~Gouv\^ea}\ \emph {et~al.}(2020)\citenamefont
  {De~Gouv\^ea}, \citenamefont {Martinez-Soler}, \citenamefont
  {Perez-Gonzalez},\ and\ \citenamefont {Sen}}]{DeGouvea:2020ang}%
  \BibitemOpen
  \bibfield  {author} {\bibinfo {author} {\bibfnamefont {A.}~\bibnamefont
  {De~Gouv\^ea}}, \bibinfo {author} {\bibfnamefont {I.}~\bibnamefont
  {Martinez-Soler}}, \bibinfo {author} {\bibfnamefont {Y.~F.}\ \bibnamefont
  {Perez-Gonzalez}}, \ and\ \bibinfo {author} {\bibfnamefont {M.}~\bibnamefont
  {Sen}},\ }\href {\doibase 10.1103/PhysRevD.102.123012} {\bibfield  {journal}
  {\bibinfo  {journal} {Phys. Rev. D}\ }\textbf {\bibinfo {volume} {102}},\
  \bibinfo {pages} {123012} (\bibinfo {year} {2020})},\ \Eprint
  {http://arxiv.org/abs/2007.13748} {arXiv:2007.13748 [hep-ph]} \BibitemShut
  {NoStop}%
\bibitem [{\citenamefont {Rink}\ and\ \citenamefont
  {Sen}(2022)}]{Rink:2022nvw}%
  \BibitemOpen
  \bibfield  {author} {\bibinfo {author} {\bibfnamefont {T.}~\bibnamefont
  {Rink}}\ and\ \bibinfo {author} {\bibfnamefont {M.}~\bibnamefont {Sen}},\
  }\href@noop {} {\  (\bibinfo {year} {2022})},\ \Eprint
  {http://arxiv.org/abs/2211.16520} {arXiv:2211.16520 [hep-ph]} \BibitemShut
  {NoStop}%
\bibitem [{\citenamefont {An}\ \emph {et~al.}(2016)\citenamefont {An} \emph
  {et~al.}}]{JUNO:2015zny}%
  \BibitemOpen
  \bibfield  {author} {\bibinfo {author} {\bibfnamefont {F.}~\bibnamefont {An}}
  \emph {et~al.} (\bibinfo {collaboration} {JUNO}),\ }\href {\doibase
  10.1088/0954-3899/43/3/030401} {\bibfield  {journal} {\bibinfo  {journal} {J.
  Phys. G}\ }\textbf {\bibinfo {volume} {43}},\ \bibinfo {pages} {030401}
  (\bibinfo {year} {2016})},\ \Eprint {http://arxiv.org/abs/1507.05613}
  {arXiv:1507.05613 [physics.ins-det]} \BibitemShut {NoStop}%
\bibitem [{\citenamefont {Abusleme}\ \emph {et~al.}(2023)\citenamefont
  {Abusleme} \emph {et~al.}}]{JUNO:2023zty}%
  \BibitemOpen
  \bibfield  {author} {\bibinfo {author} {\bibfnamefont {A.}~\bibnamefont
  {Abusleme}} \emph {et~al.} (\bibinfo {collaboration} {JUNO}),\ }\href@noop {}
  {\  (\bibinfo {year} {2023})},\ \Eprint {http://arxiv.org/abs/2303.03910}
  {arXiv:2303.03910 [hep-ex]} \BibitemShut {NoStop}%
\bibitem [{\citenamefont {Anamiati}\ \emph {et~al.}(2019)\citenamefont
  {Anamiati}, \citenamefont {De~Romeri}, \citenamefont {Hirsch}, \citenamefont
  {Ternes},\ and\ \citenamefont {T\'ortola}}]{Anamiati:2019maf}%
  \BibitemOpen
  \bibfield  {author} {\bibinfo {author} {\bibfnamefont {G.}~\bibnamefont
  {Anamiati}}, \bibinfo {author} {\bibfnamefont {V.}~\bibnamefont {De~Romeri}},
  \bibinfo {author} {\bibfnamefont {M.}~\bibnamefont {Hirsch}}, \bibinfo
  {author} {\bibfnamefont {C.~A.}\ \bibnamefont {Ternes}}, \ and\ \bibinfo
  {author} {\bibfnamefont {M.}~\bibnamefont {T\'ortola}},\ }\href {\doibase
  10.1103/PhysRevD.100.035032} {\bibfield  {journal} {\bibinfo  {journal}
  {Phys. Rev. D}\ }\textbf {\bibinfo {volume} {100}},\ \bibinfo {pages}
  {035032} (\bibinfo {year} {2019})},\ \Eprint
  {http://arxiv.org/abs/1907.00980} {arXiv:1907.00980 [hep-ph]} \BibitemShut
  {NoStop}%
\bibitem [{\citenamefont {Kobayashi}\ and\ \citenamefont
  {Lim}(2001)}]{Kobayashi:2000md}%
  \BibitemOpen
  \bibfield  {author} {\bibinfo {author} {\bibfnamefont {M.}~\bibnamefont
  {Kobayashi}}\ and\ \bibinfo {author} {\bibfnamefont {C.~S.}\ \bibnamefont
  {Lim}},\ }\href {\doibase 10.1103/PhysRevD.64.013003} {\bibfield  {journal}
  {\bibinfo  {journal} {Phys. Rev. D}\ }\textbf {\bibinfo {volume} {64}},\
  \bibinfo {pages} {013003} (\bibinfo {year} {2001})},\ \Eprint
  {http://arxiv.org/abs/hep-ph/0012266} {arXiv:hep-ph/0012266} \BibitemShut
  {NoStop}%
\bibitem [{\citenamefont {Parke}(1986)}]{Parke:1986jy}%
  \BibitemOpen
  \bibfield  {author} {\bibinfo {author} {\bibfnamefont {S.~J.}\ \bibnamefont
  {Parke}},\ }\href {\doibase 10.1103/PhysRevLett.57.1275} {\bibfield
  {journal} {\bibinfo  {journal} {Phys. Rev. Lett.}\ }\textbf {\bibinfo
  {volume} {57}},\ \bibinfo {pages} {1275} (\bibinfo {year} {1986})},\ \Eprint
  {http://arxiv.org/abs/2212.06978} {arXiv:2212.06978 [hep-ph]} \BibitemShut
  {NoStop}%
\bibitem [{\citenamefont {de~Gouvea}\ \emph {et~al.}(1999)\citenamefont
  {de~Gouvea}, \citenamefont {Friedland},\ and\ \citenamefont
  {Murayama}}]{deGouvea:1999wg}%
  \BibitemOpen
  \bibfield  {author} {\bibinfo {author} {\bibfnamefont {A.}~\bibnamefont
  {de~Gouvea}}, \bibinfo {author} {\bibfnamefont {A.}~\bibnamefont
  {Friedland}}, \ and\ \bibinfo {author} {\bibfnamefont {H.}~\bibnamefont
  {Murayama}},\ }\href {\doibase 10.1103/PhysRevD.60.093011} {\bibfield
  {journal} {\bibinfo  {journal} {Phys. Rev. D}\ }\textbf {\bibinfo {volume}
  {60}},\ \bibinfo {pages} {093011} (\bibinfo {year} {1999})},\ \Eprint
  {http://arxiv.org/abs/hep-ph/9904399} {arXiv:hep-ph/9904399} \BibitemShut
  {NoStop}%
\bibitem [{\citenamefont {Friedland}(2000)}]{Friedland:2000cp}%
  \BibitemOpen
  \bibfield  {author} {\bibinfo {author} {\bibfnamefont {A.}~\bibnamefont
  {Friedland}},\ }\href {\doibase 10.1103/PhysRevLett.85.936} {\bibfield
  {journal} {\bibinfo  {journal} {Phys. Rev. Lett.}\ }\textbf {\bibinfo
  {volume} {85}},\ \bibinfo {pages} {936} (\bibinfo {year} {2000})},\ \Eprint
  {http://arxiv.org/abs/hep-ph/0002063} {arXiv:hep-ph/0002063} \BibitemShut
  {NoStop}%
\bibitem [{\citenamefont {Vinyoles}\ \emph {et~al.}(2017)\citenamefont
  {Vinyoles}, \citenamefont {Serenelli}, \citenamefont {Villante},
  \citenamefont {Basu}, \citenamefont {Bergström}, \citenamefont
  {Gonzalez-Garcia}, \citenamefont {Maltoni}, \citenamefont {Peña-Garay},\
  and\ \citenamefont {Song}}]{Vinyoles:2017}%
  \BibitemOpen
  \bibfield  {author} {\bibinfo {author} {\bibfnamefont {N.}~\bibnamefont
  {Vinyoles}}, \bibinfo {author} {\bibfnamefont {A.~M.}\ \bibnamefont
  {Serenelli}}, \bibinfo {author} {\bibfnamefont {F.~L.}\ \bibnamefont
  {Villante}}, \bibinfo {author} {\bibfnamefont {S.}~\bibnamefont {Basu}},
  \bibinfo {author} {\bibfnamefont {J.}~\bibnamefont {Bergström}}, \bibinfo
  {author} {\bibfnamefont {M.~C.}\ \bibnamefont {Gonzalez-Garcia}}, \bibinfo
  {author} {\bibfnamefont {M.}~\bibnamefont {Maltoni}}, \bibinfo {author}
  {\bibfnamefont {C.}~\bibnamefont {Peña-Garay}}, \ and\ \bibinfo {author}
  {\bibfnamefont {N.}~\bibnamefont {Song}},\ }\href {\doibase
  10.3847/1538-4357/835/2/202} {\bibfield  {journal} {\bibinfo  {journal} {The
  Astrophysical Journal}\ }\textbf {\bibinfo {volume} {835}},\ \bibinfo {pages}
  {202} (\bibinfo {year} {2017})}\BibitemShut {NoStop}%
\bibitem [{\citenamefont {Giunti}\ and\ \citenamefont {Kim}(2007)}]{GiuntiKim}%
  \BibitemOpen
  \bibfield  {author} {\bibinfo {author} {\bibfnamefont {C.}~\bibnamefont
  {Giunti}}\ and\ \bibinfo {author} {\bibfnamefont {C.~W.}\ \bibnamefont
  {Kim}},\ }\href {\doibase 10.1093/acprof:oso/9780198508717.001.0001} {\emph
  {\bibinfo {title} {{Fundamentals of Neutrino Physics and Astrophysics}}}}\
  (\bibinfo  {publisher} {Oxford University Press},\ \bibinfo {year}
  {2007})\BibitemShut {NoStop}%
\bibitem [{\citenamefont {Lopes}\ and\ \citenamefont
  {Turck-Chièze}(2013)}]{Lopes:2013}%
  \BibitemOpen
  \bibfield  {author} {\bibinfo {author} {\bibfnamefont {I.}~\bibnamefont
  {Lopes}}\ and\ \bibinfo {author} {\bibfnamefont {S.}~\bibnamefont
  {Turck-Chièze}},\ }\href {\doibase 10.1088/0004-637X/765/1/14} {\bibfield
  {journal} {\bibinfo  {journal} {The Astrophysical Journal}\ }\textbf
  {\bibinfo {volume} {765}},\ \bibinfo {pages} {14} (\bibinfo {year}
  {2013})}\BibitemShut {NoStop}%
\bibitem [{\citenamefont {Aalbers}\ \emph {et~al.}(2020)\citenamefont {Aalbers}
  \emph {et~al.}}]{Aalbers:2020}%
  \BibitemOpen
  \bibfield  {author} {\bibinfo {author} {\bibfnamefont {J.}~\bibnamefont
  {Aalbers}} \emph {et~al.} (\bibinfo {collaboration} {DARWIN}),\ }\href
  {\doibase 10.1140/epjc/s10052-020-08602-7} {\bibfield  {journal} {\bibinfo
  {journal} {The European Physical Journal C}\ }\textbf {\bibinfo {volume}
  {80}} (\bibinfo {year} {2020}),\ 10.1140/epjc/s10052-020-08602-7}\BibitemShut
  {NoStop}%
\bibitem [{\citenamefont {Abusleme}\ \emph {et~al.}(2021)\citenamefont
  {Abusleme} \emph {et~al.}}]{JUNO:2021kxb}%
  \BibitemOpen
  \bibfield  {author} {\bibinfo {author} {\bibfnamefont {A.}~\bibnamefont
  {Abusleme}} \emph {et~al.} (\bibinfo {collaboration} {JUNO}),\ }\href
  {\doibase 10.1007/JHEP11(2021)102} {\bibfield  {journal} {\bibinfo  {journal}
  {JHEP}\ }\textbf {\bibinfo {volume} {11}},\ \bibinfo {pages} {102} (\bibinfo
  {year} {2021})},\ \Eprint {http://arxiv.org/abs/2107.03669} {arXiv:2107.03669
  [physics.ins-det]} \BibitemShut {NoStop}%
\bibitem [{\citenamefont {Esteban}\ \emph {et~al.}(2020)\citenamefont
  {Esteban}, \citenamefont {Gonzalez-Garcia}, \citenamefont {Maltoni},
  \citenamefont {Schwetz},\ and\ \citenamefont {Zhou}}]{Esteban:2020cvm}%
  \BibitemOpen
  \bibfield  {author} {\bibinfo {author} {\bibfnamefont {I.}~\bibnamefont
  {Esteban}}, \bibinfo {author} {\bibfnamefont {M.~C.}\ \bibnamefont
  {Gonzalez-Garcia}}, \bibinfo {author} {\bibfnamefont {M.}~\bibnamefont
  {Maltoni}}, \bibinfo {author} {\bibfnamefont {T.}~\bibnamefont {Schwetz}}, \
  and\ \bibinfo {author} {\bibfnamefont {A.}~\bibnamefont {Zhou}},\ }\href
  {\doibase 10.1007/JHEP09(2020)178} {\bibfield  {journal} {\bibinfo  {journal}
  {JHEP}\ }\textbf {\bibinfo {volume} {09}},\ \bibinfo {pages} {178} (\bibinfo
  {year} {2020})},\ \Eprint {http://arxiv.org/abs/2007.14792} {arXiv:2007.14792
  [hep-ph]} \BibitemShut {NoStop}%
\bibitem [{\citenamefont {Aalbers}\ \emph {et~al.}(2016)\citenamefont {Aalbers}
  \emph {et~al.}}]{DARWIN:2016hyl}%
  \BibitemOpen
  \bibfield  {author} {\bibinfo {author} {\bibfnamefont {J.}~\bibnamefont
  {Aalbers}} \emph {et~al.} (\bibinfo {collaboration} {DARWIN}),\ }\href
  {\doibase 10.1088/1475-7516/2016/11/017} {\bibfield  {journal} {\bibinfo
  {journal} {JCAP}\ }\textbf {\bibinfo {volume} {11}},\ \bibinfo {pages} {017}
  (\bibinfo {year} {2016})},\ \Eprint {http://arxiv.org/abs/1606.07001}
  {arXiv:1606.07001 [astro-ph.IM]} \BibitemShut {NoStop}%
\end{thebibliography}%
	
\end{document}